\documentclass[preprint]{aastex63}
\usepackage{amsmath}
\usepackage{enumitem}
\begin{document}
\title{Structure and Internal Kinematics of Nine Inner Milky Way Globular Clusters\footnote{Based on
    observations made with the NASA/ESA Hubble Space Telescope, obtained at the
    Space Telescope Science Institute, which is operated by the Association of
    Universities for Research in Astronomy, Inc., under NASA contract NAS
    5-26555.  These observations are associated with program GO-15065.}} 

\correspondingauthor{Roger E. Cohen}
\email{rcohen@stsci.edu}

\author{Roger E. Cohen}
\affiliation{Space Telescope Science Institute, 3700 San Martin
  Drive, Baltimore, MD 21218, USA}

\author{Andrea Bellini}
\affiliation{Space Telescope Science Institute, 3700 San Martin
  Drive, Baltimore, MD 21218, USA}

\author{Mattia Libralato}
\affiliation{Space Telescope Science Institute, 3700 San Martin Drive, Baltimore, MD 21218, USA}

\author{Matteo Correnti}
\affiliation{Space Telescope Science Institute, 3700 San Martin Drive, Baltimore, MD 21218, USA}

\author{Thomas M. Brown}
\affiliation{Space Telescope Science Institute, 3700 San Martin
  Drive, Baltimore, MD 21218, USA}

\author{Jason S. Kalirai}
\affiliation{Johns Hopkins University Applied Physics Laboratory, 11101 Johns Hopkins Road, Laurel, MD 21723, USA}

\begin{abstract}

This study constitutes part of a larger effort aimed at better characterizing the Galactic globular clusters (GGCs) located
towards the inner Milky Way bulge and disk.  Here, we focus on internal kinematics of nine GGCs,
obtained from space-based imaging over time baselines of $>$9 years.  We exploit multiple avenues to assess the dynamical state of the target GGCs, constructing radial profiles of projected stellar density, proper motion dispersion, and anisotropy.  We posit that two-thirds (6/9) of our target GGCs are in an advanced dynamical state, and are close to (or have recently undergone) core collapse, supported by at least two lines of evidence: First, we find relatively steep proper motion dispersion profiles, in accord with literature values for core-collapsed GGCs.  Second, we find that our sample is, in the mean, isotropic even out to their half-light radii, although one of our target clusters (NGC 6380) is tangentially anisotropic at $>$1$\sigma$ beyond its half-light radius, in accord with theoretical predictions for clusters evolving in strong tidal fields.  Our proper motion dispersion and anisotropy profiles are made publicly available. 

\end{abstract}

\section{Introduction} \label{sec:intro}

The internal kinematics of Galactic globular clusters (GGCs) provides a valuable window into their dynamical state.  At the same time, clusters serve as critical testbeds for comparison with simulations of cluster dynamical evolution.  In particular, by combining cluster structural parameters with radial profiles of proper motion dispersion and anisotropy, \citet{watkinskin} were able to characterize a relationship between anisotropy and relaxation time, showing that dynamically older clusters were isotropic out to larger radii.  They also found that more concentrated clusters exhibited steeper slopes in their dispersion profiles.  This study has the advantage that it addresses a statistically significant sample of 22 GGCs with a range of dynamical ages, providing a solid empirical footing to which other samples may be compared.

Observational characterization of the internal kinematics of 
GGCs also provides important constraints on theoretical predictions.  
Specifically, simulated GGCs never reach full energy equipartion \citep{bm03,trentivdm,bianchini16,newmodels}, largely consistent with measurements of internal kinematics over a useful baseline in stellar mass within GGCs \citep{bellini_pm,libralato_pm,libralato6352,watkins20}.  In addition, the observed behavior of GGC radial anisotropy profiles \citep{watkinskin,bellini_47tuc,libralato6352,raso1261} is consistent with model predictions that tidally underfilling clusters generally relax from the inside out \citep[e.g.][]{tiongco16a,zocchi16}.

Such data-model comparisons are feasible 
largely because the observations consist of 
proper motions rather than line-of-sight velocity measurements.  While the availability of both is ideal \citep[i.e.][]{watkins_dyndist,bellini_47tuc}, proper motions have several advantages: First, they describe stellar motions in two orthogonal components, allowing for a measurement of anisotropy.  Second, they generally yield larger sample sizes, in many cases over a significant baseline of stellar mass.  Third, because proper motions result from the comparison of entire images rather than targeting individual sources a priori, information concerning the relative motions of field stars and background galaxies is obtained as well.

Despite their advantages, the use of relative proper motions for GGC internal kinematics requires exquisite photometric and astrometric precision, ideally in more than one observational epoch, which became feasible only recently.  
For this reason, the GGCs located towards the inner Milky Way bulge and disk have, until now, been excluded from any analysis of internal kinematics; the necessary deep, multi-epoch imaging simply did not exist.  Furthermore, in the rare cases where even single-epoch deep imaging has been analyzed, GGCs towards the bulge turn out to have structural parameters differing significantly from their catalog values based on optical integrated light \citep[e.g.][]{lanzoni_ter5dens,cohen_6544,saracino_liller1}.  Meanwhile, existing compilations, taken at face value, suggest that inner Milky Way GGCs are, from a dynamical point of view, a particularly interesting subset: They are preferentially concentrated, with a high incidence of core-collapse
(\citealp*{h96}, 2010 edition, hereafter \citeauthor{h96}; \citealt{b19}).  Therefore, inner Milky Way GGCs represent a valuable opportunity to build up a 
sample of dynamically evolved clusters in a strong tidal field.  From the standpoint of internal kinematics and structure, such a sample 
is ideal not only for empirical comparison with a large ensemble of well-studied GGCs \citep{watkinskin}, but also with theoretical predictions regarding clusters evolving in tidal fields \citep{bm03,vesperini14,sollima15,bianchini17}.

With this in mind, we exploit our recent second-epoch \textit{Hubble Space Telescope} (\textit{HST}) imaging of nine GGCs located towards the inner Milky Way (GO-15065, PI:Cohen) to study the internal kinematics of these clusters.  This imaging was obtained with the primary goal of cleaning and homogenizing cluster color-magnitude diagrams (CMDs) of clusters with extant first-epoch archival imaging to measure global cluster properties (R.~E.~Cohen et al.~2020, in prep.), but we found that the precision of the available proper motions is sufficient for an internal kinematic analysis, at least of a fixed-mass population (see Sect.~\ref{cutsect}).

The remainder of this paper is organized as follows: In the next section, we summarize our observations and data reduction strategy to yield precision astrophotometric catalogs.  In Sect.~\ref{structsect} we use single-mass radial number density profiles to measure the structural parameters of our target clusters, and in Sect.~\ref{kinsect} we present radial profiles of proper motion dispersion and anisotropy.  In Sect.~\ref{discusssect} we use our results to assess the dynamical state of our target clusters, 
and in the final section we summarize our results and discuss avenues for future investigation.

\section{Observations and Data Reduction \label{obssect}} 

We make use of multi-epoch \textit{HST} imaging of nine GGCs located towards the Milky Way bulge and inner disk.  These clusters were selected based on extant first-epoch \textit{HST} imaging of sufficient quality to enable precision astrometry faintward of their main sequence turnoffs (MSTOs).  We have obtained second-epoch photometry of all nine of these clusters with ACS/WFC onboard \textit{HST} in the F606W and F814W filters (GO-15065, PI:Cohen), facilitating multi-epoch proper motion analyses while homogenizing the photometry to the same system as the large, well-studied sample of 65 GGCs observed in the \textit{HST} GGC Treasury Survey \citep{ataggc}.  The observations we use are summarized in Table \ref{obstab} along with catalog values of the cluster heliocentric distances and iron abundances, although these should probably only be regarded as indicative (we return to this point later in Sect.~\ref{futuresect}) and are included to demonstrate the approximate range of values spanned by the target clusters.  The multi-epoch imaging we use covers time baselines from 9.1 to 16.0 yr, sufficient not only for cluster-field separation based on relative proper motions, but also for analysis of the \textit{internal} kinematics of our target clusters, which is the focus of the present study.

The photometric and astrometric reduction techniques we use have been presented in detail elsewhere \citep{bellini14,bellini_wcen_phot,bellini_pm,libralato_pm}, and a detailed description of the observations, proper-motion-cleaned CMDs and differential reddening corrections for our target clusters is presented in a companion paper (Cohen et al.~in prep).  Therefore, we summarize them very briefly here for convenience.

For each target cluster, point spread function fitting (PSF) photometry is performed separately for each epoch, in multiple passes, as described in \citet{bellini_wcen_phot} and \citet{bellini_pm}, and the resulting instrumental catalogs are calibrated to the Vegamag system \citep[e.g.][]{bellini_wcen_phot,nardiellocats,libralato6352}.  Next, per-image distortion-corrected positions are used to calculate individual stellar proper motions relative to the bulk cluster motion using a least-squares fit of a straight line to the position of each star versus time in each coordinate.  This procedure is performed iteratively, and the resulting proper motions are corrected for any remaining systematic residuals as a function of color, magnitude, and position, propagating in quadrature the uncertainties on these corrections \citep{bellini14,bellini_pm,libralato_pm}.  The photometry is corrected for differential reddening, also iteratively, by shifting each star along the reddening vector to place it on an empirically determined fiducial sequence, by an amount determined using local, well-measured proper motion members \citep[e.g.][]{milone12,bellini_wcen_dr}.

We make several cuts in our astrophotometric catalogs to retain only well-measured stellar sources 
(note that an additional set of more stringent cuts are necessary for internal kinematics, described in Sect.~\ref{cutsect}).  These cuts are applied simultaneously (rather than sequentially) on a per-filter, per-epoch basis using various photometric diagnostic parameters as follows:
\begin{enumerate}
\item The fractional flux within the PSF fitting radius from neighbors (prior to neighbor subtraction) is denoted $o$.
We set a fixed upper limit on $o$ on a per-filter, per-cluster basis by retaining only the stars with the 70\% of smallest $o$ values.  In practice, this resulted in maximum values of $o<$0.7 in all cases, a slightly harsher cut than other studies which typically make cuts at a fixed value of $o<$1 \citep[e.g.][]{bellini_wcen_dr}.  However, this turns out to be inconsequential to our results, and the impact of our $o$ cut is similar across our sample of target clusters.  Since $o$ describes \textit{relative}, rather than absolute flux contribution from neighbors, the stars which passed our other quality cuts described below and failed only our $o$ cuts are preferentially faint, with 97\% of them in the median (and $<$90\% in all cases) lying faintward of the upper main sequence and red giant branch where we perform our analyses (see Sects.~\ref{denssect} and \ref{cutsect}). \label{firstbasicitem}

\item RADXS is the ratio between the flux outside the PSF fitting radius to the flux predicted by the model PSF, and can therefore be positive (i.e.~for galaxies) or negative (i.e.~for cosmic rays).  We retain only sources with -0.05$\leq$RADXS$\leq$0.05.

\item The QFIT value is essentially a linear correlation coefficient between observed pixel values and those predicted by the PSF model, so that a perfect fit will have QFIT=1 and worse PSF fits will have 0$<$QFIT$<$1.  Because QFIT becomes strongly magnitude-dependent towards fainter magnitudes (e.g.~fig.~8 of \citealt{bellini_wcen_phot}), we exclude stars with the lowest 7.5\% of QFIT values at their magnitude, and exclude all stars with QFIT$<$0.7 regardless of magnitude.  As with the cut on $o$, the impact of the QFIT cut does not appreciably vary over our target cluster sample since we intentionally kept the observing strategy nearly identical for all of the second-epoch imaging we use here. \label{lastbasicitem}
\end{enumerate}

\begin{figure}
\gridline{\fig{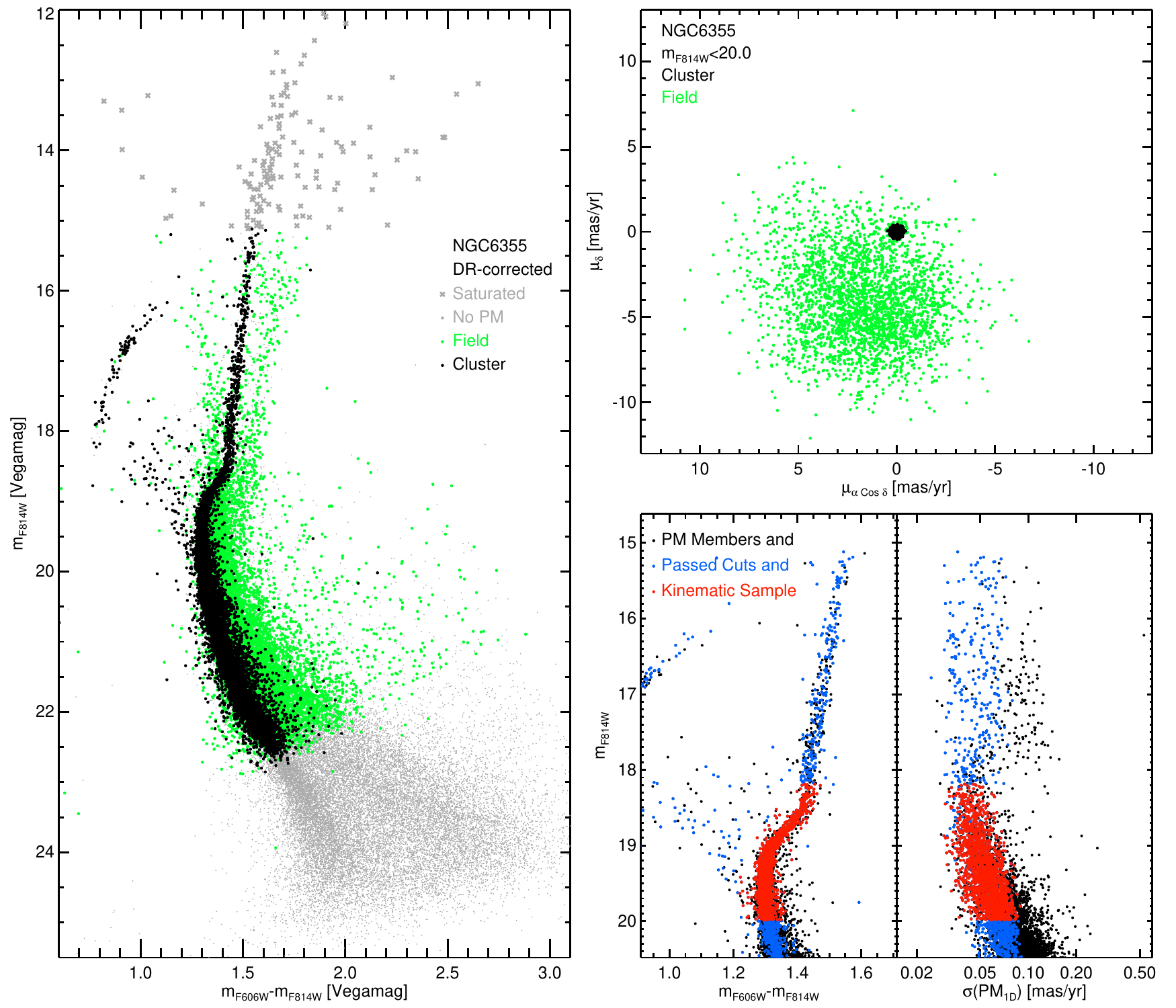}{0.95\textwidth}{}}
\caption{Illustration of stars passing our quality cuts, for the example case of NGC 6355.  \textbf{(Left:)} CMD, corrected for differential reddening, showing all stars passing the basic photometric quality cuts described in Sect.~\ref{obssect}, except for saturated stars, which are shown as grey crosses.  Stars without valid proper motions are shown in grey, whereas proper-motion-selected cluster and field stars are shown using black and green circles respectively.  \textbf{(Upper Right:)} Vector point diagram showing relative proper motions of cluster and field stars.  \textbf{(Lower Right:)} CMD and proper motion uncertainties of cluster stars passing the basic cuts listed in Sect.~\ref{obssect} (black), zoomed in on the upper main sequence and red giant branch, to highlight the subset of these that pass the more stringent photometric and astrometric quality cuts for internal kinematics described in Sect.~\ref{cutsect} (blue points), and the mono-mass sample selected within this subset for kinematic analysis based on their proper motion uncertainties (red points). \label{kinexamplefig}}
\end{figure}

In the left panel of Fig.~\ref{kinexamplefig}, we show an example differential-reddening-corrected CMD from our second-epoch imaging of NGC 6355.  This CMD shows all of the sources that passed our quality cuts, with the exception of saturated stars, which are shown as grey crosses and excluded from any further analysis here.  Of the sources passing our quality cuts, those without measured proper motions are shown as grey points, most of which lie at the faint end of the CMD because our second epoch imaging is, by design, deeper than the first epoch archival imaging.  The upper and lower right-hand panels of Fig.~\ref{kinexamplefig} refer to additional more stringent cuts made to select a sample for kinematic analysis (see Sect.~\ref{cutsect}).

We perform artificial star tests to quantify photometric incompleteness as a function of color, magnitude, and distance from the cluster center (i.e.~crowding) in our second epoch imaging.  For each cluster, 10$^{6}$ artificial stars are inserted, one at a time, with a realistic luminosity function (i.e.~exponentially increasing towards fainter magnitudes) and spatial distribution (more crowded towards the cluster center, based on an exponential radial density profile), and given a color distribution based on the observed CMD.  
Artificial star tests are performed using software based on that used for the ACS GGC Treasury Survey \citep{andersonggc}, and artificial stars are considered recovered if they are found within 0.5 pix of their input positions and pass all of the quality cuts listed in items \ref{firstbasicitem}-\ref{lastbasicitem} above.

\begin{deluxetable}{lccccccc}
\tabletypesize{\small} %scriptsize
\tablecaption{Summary of Observations \label{obstab}}
\tablehead{
\colhead{Cluster} & \colhead{$R_{\sun}$\tablenotemark{a}} & \colhead{[Fe/H]\tablenotemark{b}} & \colhead{$t_{base}$} & \colhead{Epoch} & \colhead{GO} & \colhead{Filter} & \colhead{Exposures} \\ \colhead{} & \colhead{kpc} & \colhead{dex} &  \colhead{yr} & \colhead{} & \colhead{} & \colhead{} & \colhead{}
}
%%\colnumbers
\startdata
NGC 6256 & 9.1 & -0.62$\pm$0.09 & 9.871 & Aug 2009 & 11628 & WFC3/UVIS F555W & 3$\times$360s \\
	 &   & &      &          & 11628 & WFC3/UVIS F814W & 3$\times$100s \\
	 &   & &     & Jun 2019 & 15065 & ACS/WFC F606W & 1$\times$60s $+$ 4$\times$502s \\
	 &   & &     &          & 15065 & ACS/WFC F814W & 1$\times$60s $+$ 4$\times$505s \\ 
NGC 6325 & 7.8 & -1.37$\pm$0.14 & 9.131 & May 2010 & 11628 & WFC3/UVIS F438W & 3$\times$435s \\
	 &   & &     &          & 11628 & WFC3/UVIS F555W & 3$\times$85s \\
	 &   & &     & Jun 2019 & 15065 & ACS/WFC F606W & 1$\times$30s $+$ 4$\times$494s \\
	 &   & &    &          & 15065 & ACS/WFC F814W & 1$\times$30s $+$ 4$\times$493s \\
NGC 6342 & 8.4 & -0.49$\pm$0.14 & 9.871 & Aug 2009 & 11628 & WFC3/UVIS F438W & 3$\times$420s \\
	 &   & &     &          & 11628 & WFC3/UVIS F555W & 3$\times$80s \\
	 &   & &     & Jun 2019 & 15065 & ACS/WFC F606W & 1$\times$10s $+$ 4$\times$493s \\
	 &   & &     &          & 15065 & ACS/WFC F814W & 1$\times$10s $+$ 4$\times$493s \\
NGC 6355 & 9.0 & -1.33$\pm$0.14 & 9.881 & Aug 2009 & 11628 & WFC3/UVIS F438W & 3$\times$440s \\
	 &   & &     &          & 11628 & WFC3/UVIS F555W & 3$\times$80s \\
	 &   & &     & Jun 2019 & 15065 & ACS/WFC F606W & 1$\times$30s $+$ 4$\times$495s \\
	 &   & &     &          & 15065 & ACS/WFC F814W & 1$\times$30s $+$ 4$\times$496s \\
NGC 6380 & 9.2 & -0.40$\pm$0.09 & 9.117 & Mar 2010 & 11628 & WFC3/UVIS F555W & 3$\times$440s \\
	 &   & &     &          & 11628 & WFC3/UVIS F814W & 3$\times$80s \\
	 &   & &     & Apr 2019 & 15065 & ACS/WFC F606W & 1$\times$60s $+$ 4$\times$502s \\
	 &   & &    &          & 15065 & ACS/WFC F814W & 1$\times$60s $+$ 4$\times$502s \\
NGC 6401 & 7.7 & -1.01$\pm$0.14 & 16.048 & Jul 2003 & 9799 & ACS/WFC F606W & 2$\times$340s \\
	 &   & &     &           & 9799 & ACS/WFC F814W & 2$\times$240s \\
	 &   & &     & Aug 2019 & 15065 & ACS/WFC F606W & 1$\times$60s $+$ 4$\times$487s \\
	 &   & &     &          & 15065 & ACS/WFC F814W & 1$\times$60s $+$ 4$\times$487s \\
NGC 6453 & 10.7 & -1.48$\pm$0.14 & 9.123 & May 2010 & 11628 & WFC3/UVIS F438W & 3$\times$450s \\
	 &   & &     &          & 11628 & WFC3/UVIS F555W & 3$\times$80s \\
	 &   & &     & Jun 2019 & 15065 & ACS/WFC F606W & 1$\times$40s $+$ 4$\times$498s \\
	 &   & &     &          & 15065 & ACS/WFC F814W & 1$\times$30s $+$ 4$\times$499s \\
NGC 6558 & 7.4 & -1.37$\pm$0.14 & 15.911 & Sep 2003 & 9799 & ACS/WFC F606W & 1$\times$340s \\
	 &   & &     &           & 9799 & ACS/WFC F814W & 2$\times$10s $+$ 1$\times$340s \\
	 &   & &     & Aug 2019 & 15065 & ACS/WFC F606W & 1$\times$10s $+$ 4$\times$498s \\
	 &   & &     &          & 15065 & ACS/WFC F814W & 1$\times$10s $+$ 4$\times$498s \\
NGC 6642 & 8.6 & -1.19$\pm$0.14 & 15.144 & Mar 2004 & 9799 & ACS/WFC F606W & 1$\times$10s $+$ 1$\times$340s \\
	 &   & &     &           & 9799 & ACS/WFC F814W & 1$\times$10s $+$ 1$\times$340s \\
	 &   & &     & Apr 2019 & 15065 & ACS/WFC F606W & 1$\times$10s $+$ 4$\times$493s \\
	 &   & &     &          & 15065 & ACS/WFC F814W & 1$\times$10s $+$ 3$\times$493s \\
\enddata
%%%\tablecomments{Comments}
\tablenotetext{a}{From \citet{valenti_cat1,valenti_cat2} where available, otherwise \citeauthor{h96}.}
\tablenotetext{b}{From \citet{c09}.}
\end{deluxetable}

\section{Structural Parameters \label{structsect}}

To determine the centers of our clusters and construct number density profiles, we make use of star counts
rather than integrated light.  This choice avoids systematics due to shot noise bias from a small number of very luminous stars, which has been demonstrated to yield erroneous determinations of cluster centers (i.e.~luminosity rather than gravity centers) and artifacts in radial profiles constructed from integrated light \citep[e.g.][]{calzetti_dens,avdm}.  We use our full-frame second-epoch ACS/WFC imaging (i.e.~without requiring proper motion information) for center determinations and construction of density profiles due to its larger field of view, higher signal-to-noise, and homogeneity across our target clusters compared to the heterogenous, shallower archival first-epoch imaging.

\subsection{Cluster Centers \label{censect}}

\subsubsection{Determination of Center Coordinates}

\begin{figure}
\gridline{\fig{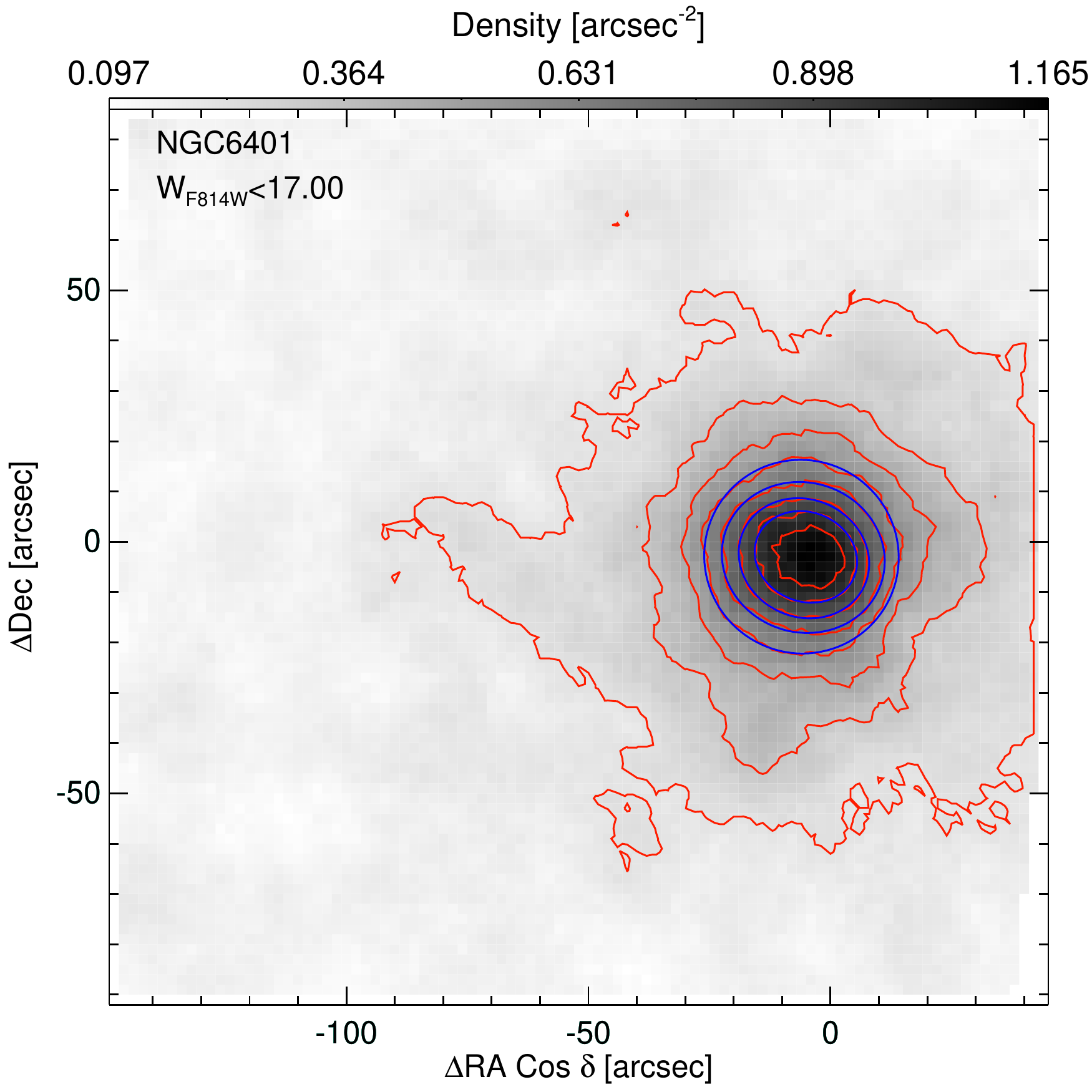}{0.6\textwidth}{}}
\caption{An example density map for NGC 6401 demonstrating our ellipse fitting procedure to determine the cluster center.  The density map (shown in greyscale according to the horizontal color bar at the top of the plot) is constructed using stars brightward of the MSTO, making a magnitude cut parallel to the reddening vector (see text for details).  Eight contours (shown in red) are fit to the density map between its minimum (non-zero) and maximum value.  Ellipses (shown in blue) are fit to the innermost four contours (excluding the central one), and the location and uncertainty of the cluster center reported in Table \ref{sbtab} is the mean and standard deviation of these individual ellipse centers. \label{cenexamplefig}} 
\end{figure}

We determine cluster centers by fitting ellipses to isodensity contours, as this method outperforms symmetry-based \textquotedblleft pie-slice\textquotedblright  methods, as discussed by \citet{goldsbury_centers} and \citet{avdm}.
To mitigate the effects of incompleteness and differential reddening, we construct a density map using only sources brighter than the MSTO in each cluster, making a magnitude cut parallel to the reddening vector\footnote{Throughout this study, we assume $(A_{F606W},A_{F814W})/E(B-V)$ = $(2.876,1.884)$ \citep{cv14}}.  Following \citet{goldsbury_centers}, we construct an oversampled density map using spatial bins placed every 2$\arcsec$, counting the number of sources within 10$\arcsec$ for each bin, and an example density map is shown in Fig.~\ref{cenexamplefig}.  Eight isodensity contours, shown in red in Fig.~\ref{cenexamplefig}, are generated from the resultant density map, evenly spaced between the minimum non-zero density and maximum density, and ellipses, shown in blue in Fig.~\ref{cenexamplefig}, are fit to each of the innermost four isodensity contours (excluding the central one).  The cluster center and its uncertainty we report, given in Table \ref{sbtab}, are then the mean and standard deviation of the individual ellipse centers.

\subsubsection{Comparison With Other Center Values}

\begin{figure}
\gridline{\fig{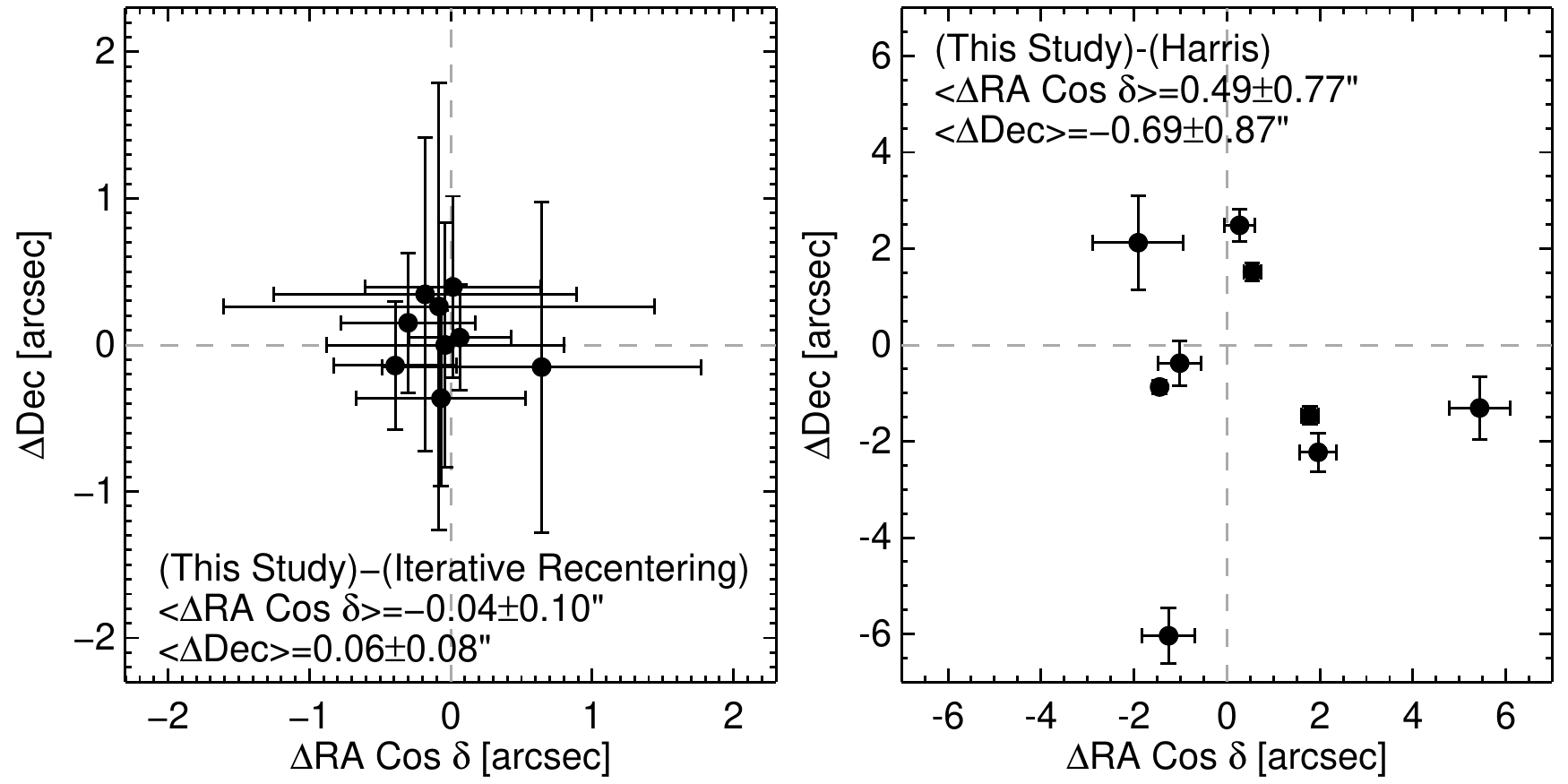}{0.8\textwidth}{}}
\caption{A comparison between the cluster center coordinates we determine using ellipse fitting to isodensity contours versus those determined by iterative recentering (left, see text) and those from the catalog of \citeauthor{h96} (right).  In each panel, the dotted grey lines correspond to a null positional offset, and the unweighted mean offset across the sample and its standard deviation are reported.  Given the agreement of our centers determined from two independent methods shown in the left-hand panel, we argue that the larger discrepancies in the right-hand panel are due to typical errors of a few arcsec in the original catalog centers, which were based on integrated light in shallow optical images (see text for details).  Note the difference in scale between the two panels.
\label{cenfig}}
\end{figure}

Given that our clusters are relatively concentrated (\citeauthor{h96}; also see Sect.~\ref{denssect}), precise measurements of their centers are crucial to obtaining reliable density profiles and constraining the structure of the cluster cores.
We first compare our centers with values given in the \citeauthor{h96} catalog, which are taken from \citet{shawlwhite} and \citet{picard} (4 clusters each) and the center of NGC 6380 is taken from \citet{djorg93}.  This comparison is shown in the right-hand panel of Fig.~\ref{cenfig}.
While there is a statistically null mean offset across the sample, the individual clusters are offset from their catalog locations typically by several arcsec.  This is not due to uncertainties in our absolute astrometry, which was calibrated to \textit{Gaia} DR2 \citep{gaiadr2} with residuals of $<$0.1$\arcsec$, nor is it due to the absolute motions of the clusters, although the latter contribution is not entirely insignificant, ranging from $\sim$0.1 to 0.4$\arcsec$ over the time since the original measurements assuming the absolute proper motions from \citet{b19}.  The offsets between our measured centers and the catalog values 
closely mirror the results of \citet{goldsbury_centers}, who used a similar technique and photometric data, and found 26/65 (40\%) of their clusters to be discrepant by more than 5$\arcsec$, while we find the same for two of our nine target clusters, also noting that they found 8/65 (12\%) to differ by more than 10$\arcsec$, while none of our target clusters show such a large offset.  Given the similarity of our results and those of \citet{goldsbury_centers}, in combination with our center uncertainties of $\leq$1$\arcsec$ (see Table \ref{sbtab}), we surmise that the original catalog centers (obtained from optical integrated light in ground-based images, in many cases with photographic plates and/or seeing worse than 1$\arcsec$) are typically incorrect by several arcsec.

To test this hypothesis, we independently redetermine our cluster centers and uncertainties using an alternative technique described and employed, for example, in \citet{montegriffo95} and \citet{lanzoni19}. This alternative technique functions via iterative recentering, in which the mean offset is calculated for each coordinate using stars inside some maximum radius $R_{max}$ from a trial center, recentering on the mean value in each of the two coordinates until convergence, indicated by a change of $<$0.01$\arcsec$ in the center location since the previous iteration.  Uncertainties are obtained by rerunning the centering procedure for different values of $R_{max}$ (from 15$\arcsec$ to 30$\arcsec$ in steps of 2.5$\arcsec$) and calculating the standard deviation of the resulting centers.  A comparison between the centers obtained via this iterative recentering method and our ellipse fitting method is shown in the left panel of Fig.~\ref{cenfig}, revealing excellent agreement: The differences in center positions are \textit{all} $<$0.7$\arcsec$ (with a median difference of $<$0.4$\arcsec$), and the mean difference in center locations (given in the left panel of Fig.~\ref{cenfig}) is not statistically significant, so we adopt the ellipse fitting centers.

For NGC 6256, our center from ellipse fitting is located 0.35$\arcsec$ from the center determined by \citet{cadelano6256} from an independent analysis of the first epoch imaging, within their quoted uncertainty of $\sim$0.4$\arcsec$.

\subsection{Radial Density Profiles \label{denssect}}
\subsubsection{Construction of Density Profiles}
We measure cluster radial density profiles by using completeness-corrected star counts, which are robust to 
biases inherent in integrated light profiles \citep[e.g.][]{calzetti_dens}.  As with the cluster center determination, we restrict our stellar sample to stars brightward of the MSTO in each cluster, and make a magnitude cut parallel to the reddening vector.  The density profile is then calculated by dividing the sample into logarithmically-spaced radial annuli\footnote{This serves as a compromise between spatial resolution and number statistics.  We also limit uncertainties by requiring that each radial bin has at least 3 observed stars per azimuthal sub-sector, which affects at most the innermost few radial bins.}, and dividing each annulus into azimuthal sub-sectors \citep[e.g.][]{djorg88,ferraro99,lanzoni19}.  We calculate the completeness-corrected density in each azimuthal sub-sector, and take the density and its uncertainty in each radial annulus as the mean and standard deviation of the densities among azimuthal sub-sectors.

Due to the size of the ACS/WFC field of view, our \textit{HST} imaging only extends to $<$200$\arcsec$ from the centers of our target clusters.  Therefore, we use star counts from \textit{Gaia} DR2 in the outer parts of our density profiles to characterize the background level \citep[e.g.][]{gaiadens,raso1261}.  Special care must be taken to mitigate incompleteness in the \textit{Gaia} photometry, so for the \textit{Gaia} catalog we also make a Wesenheit magnitude cut (parallel to the reddening vector) brightward of the cluster MSTO\footnote{Our \textit{Gaia} magnitude cut is set so that the limiting \textit{Gaia} $G$ Wesenheit magnitude is 13$\leq$$W_{G}$$\leq$14 where $W_{G}$ = $G - ((B_{P}-R_{P})(A_{G}/(A_{BP}-A_{RP})))$ assuming ($A_{BP},A_{G},A_{RP}$)/$E(B-V)$ = (3.374,2.740,2.035) \citep{cv14}}.
While we have intentionally set this magnitude cut well brightward of \textit{Gaia}'s photometric detection limit, incompleteness even at bright magnitides also varies with crowding, so we only construct the \textit{Gaia} density profile outside a minimum radius $r_{min}$ of 40-80$\arcsec$ (depending on the cluster) from the cluster center.  By combining Wesenheit magnitude cuts with radius cuts in the \textit{Gaia} sample (and no additional cuts, i.e.~on astrometric parameters, which we do not use), we are able to generate a photometric sample from \textit{Gaia} with incompleteness that does not vary with radius (note that any spatial inhomogeneity in the \textit{Gaia} catalog that persists despite our cuts will propagate to uncertainties in the density profile, since the uncertainties in each radial bin are based on the variance in the azimuthal sub-sectors).

  Importantly, by setting both the \textit{Gaia} and \textit{HST} ACS/WFC magnitude cuts brightward of the cluster MSTO, the \textit{Gaia} stellar number counts cover the same narrow cluster mass range as our ACS/WFC sample, so that we can exploit their range of radial overlap to scale the shallower wide-field (i.e.~\textit{Gaia}) portion to the \textit{HST} portion and combine them to increase our radial coverage \citep[e.g.][and references therein]{miocchi}.  Specifically, the \textit{Gaia} and ACS/WFC density profiles are compared in their range of radial overlap (between $r_{min}$ and the outer limit of the ACS/WFC field of view), and a mean scale factor (weighted by the observational uncertainties in the profile) is computed to scale the \textit{Gaia} density profile to the ACS/WFC density profile.  The fractional uncertainty of this mean scale factor is propagated to the final observed density profile, so that any mismatch in the \textit{shape} of the \textit{Gaia} and ACS/WFC density profiles in their range of radial overlap would propagate to the uncertainties in the final profile.  Fortunately, this turns out not to be a significant contributor to the error budget, as the uncertainties in the scale factor are generally quite small, with a median contribution of 5.4\%.

Lastly, after scaling the \textit{Gaia} portion of the density profile to the ACS/WFC portion, the background, calculated as the weighted mean of the densitity in the outermost (flat) portion of the density profile, is subtracted (propagating in quadrature the resulting background uncertainty).  As an external check on the structural parameters we calculate using \textit{Gaia} photometry for the outer portion of the density profile, we also construct and fit (see Sect.~\ref{densfitsect} below) density profiles using $JK_{s}$ near-IR PSF photometry from the \textit{Vista Variables in the Via Lactea} survey \citep{javier_psf}, also using radius and Wesenheit magnitude cuts, for the clusters in common, and obtain structural parameters that agree with our \textit{Gaia}-based results to within their uncertainties.  In Fig.~\ref{sbfig} we show density profiles of each cluster before background subtraction in cyan, with the background and its $\pm$1$\sigma$ uncertainty indicated using solid and dashed horizontal black lines respectively.  The final background-subtracted density profile is shown using black filled circles.

\subsubsection{Density Profile Fitting \label{densfitsect}}

We fit \citet{king66} profiles, generated with the publicly available \texttt{LIMEPY} software package \citep{limepy}\footnote{\url{https://github.com/mgieles/limepy}}, to our projected density profiles to obtain the best-fitting structural parameters.  The best-fitting model parameters and their uncertainties are
obtained using a maximum likelihood approach, and the use of flat priors renders the posterior probability distribution function (PDF) proportional to the likelihood, calculated as:

\begin{equation}
\mathcal{L} \simeq \rm{exp} \left( -\frac{1}{2}\chi^{2}\right)
\label{likeeq}
\end{equation}

Where:

\begin{equation}
\chi^{2} = \sum_{i}^{N} \frac{(\Delta s_{i})^{2}}{\sigma_{i}^{2}}
\label{chisq}
\end{equation}

Here, $\Delta s_{i}$ represents the difference in observed and model-predicted density at each of $N$ points in the observed profile, and $\sigma_{i}$ represents the uncertainty on the observed density.  We maximize the log likelihood using the \texttt{emcee} affine-invariant Markov chain monte carlo (MCMC) sampler \citep{emcee}, using 25 walkers over 300 burn-in iterations, followed by 1500 production iterations (sufficient given autocorrelation times of $<$25 iterations).  When fitting the King profiles, there are only two free parameters, namely the shape parameter $W_{0}$ (which has a one-to-one relationship with the concentration parameter $c$) and the radial scale, which can be characterized by any one of either the King radius $r_{\rm 0}$, the half-light and half-mass radii $r_{\rm hl}$ and $r_{\rm hm}$, the observational core (i.e.~half-power) radius $r_{\rm c,obs}$, or the tidal radius $r_{\rm t}$ since all of these radii have a fixed relationship to each other at a given $W_{0}$ (as discussed by e.g.~\citealt{mvdm,chatterjee_cc}).  The best-fit values (the median over all post-burnin MCMC iterations) and their uncertainties (16th and 84th percentiles) for these structural parameters are listed for our target clusters in Table \ref{sbtab}.

\begin{figure}
\gridline{\fig{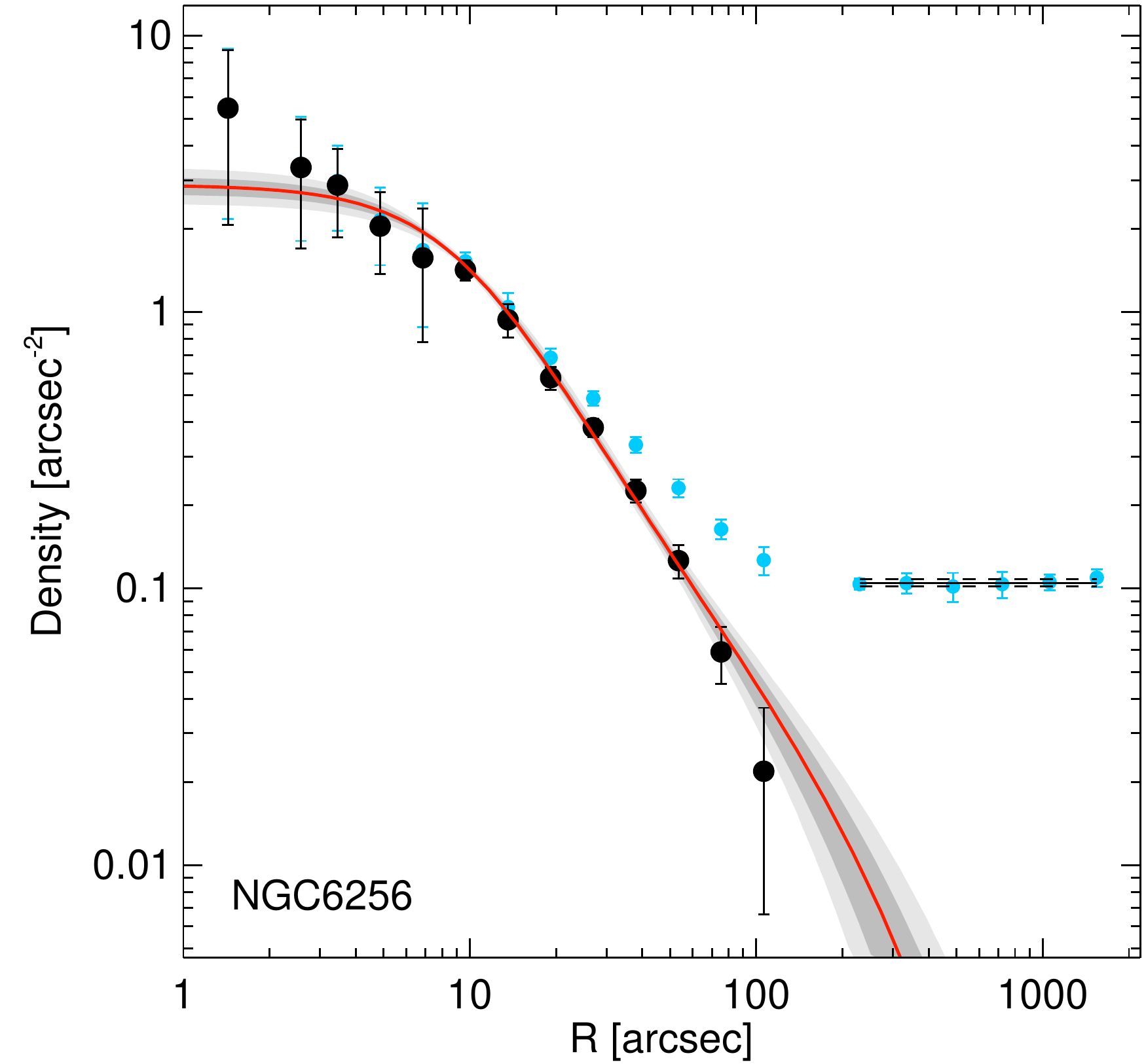}{0.33\textwidth}{}
	  \fig{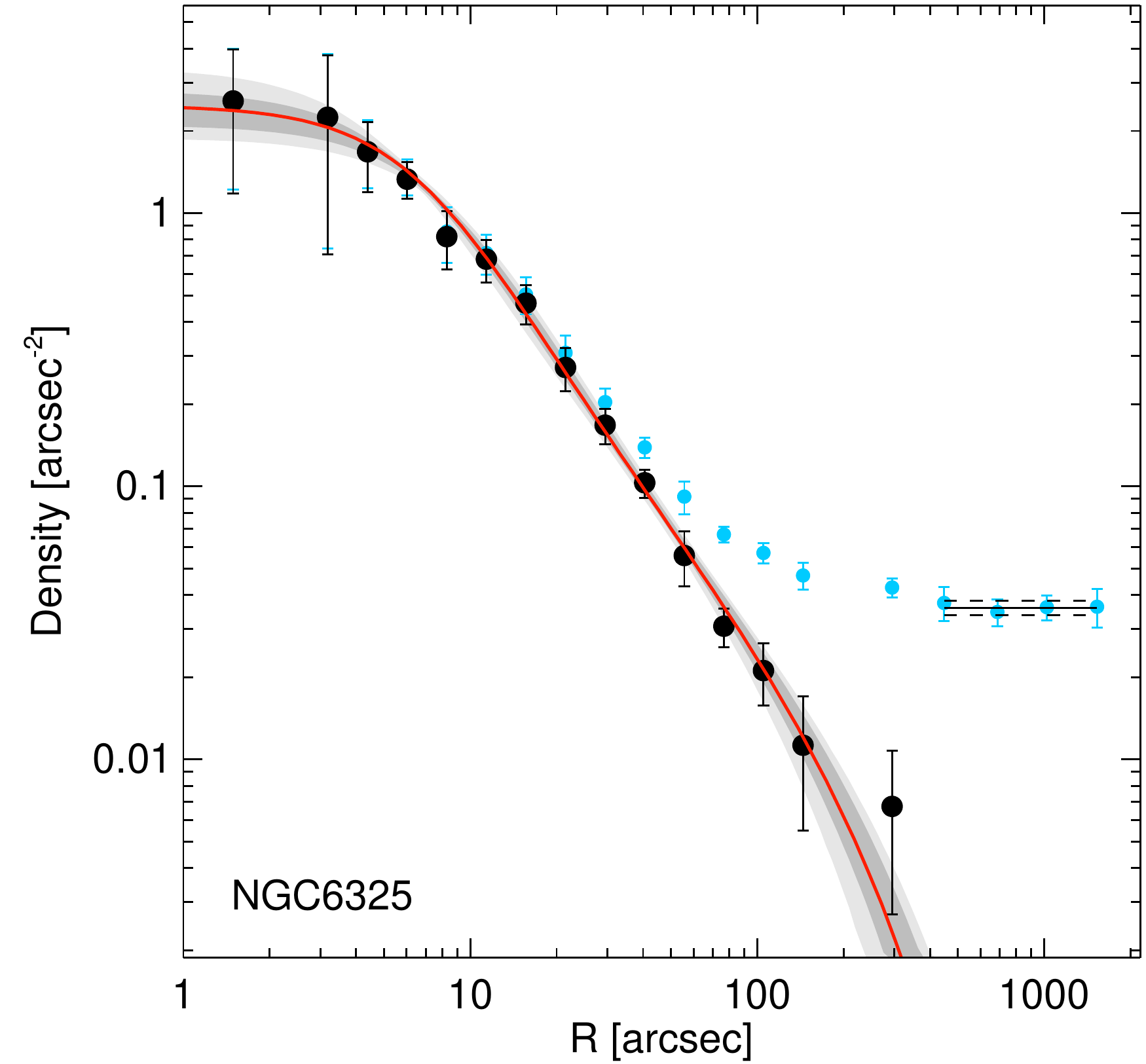}{0.33\textwidth}{}
          \fig{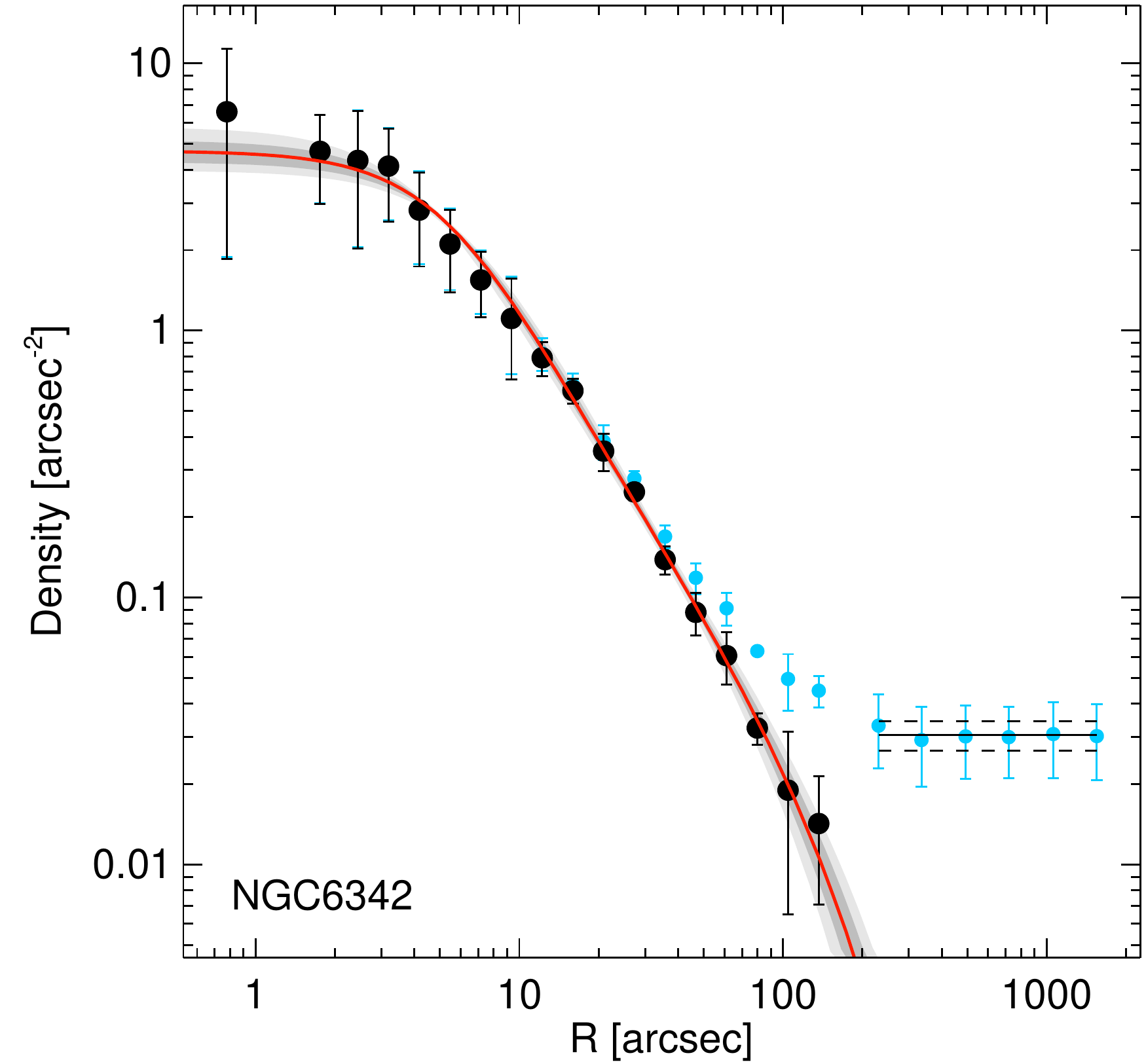}{0.33\textwidth}{}}
\gridline{\fig{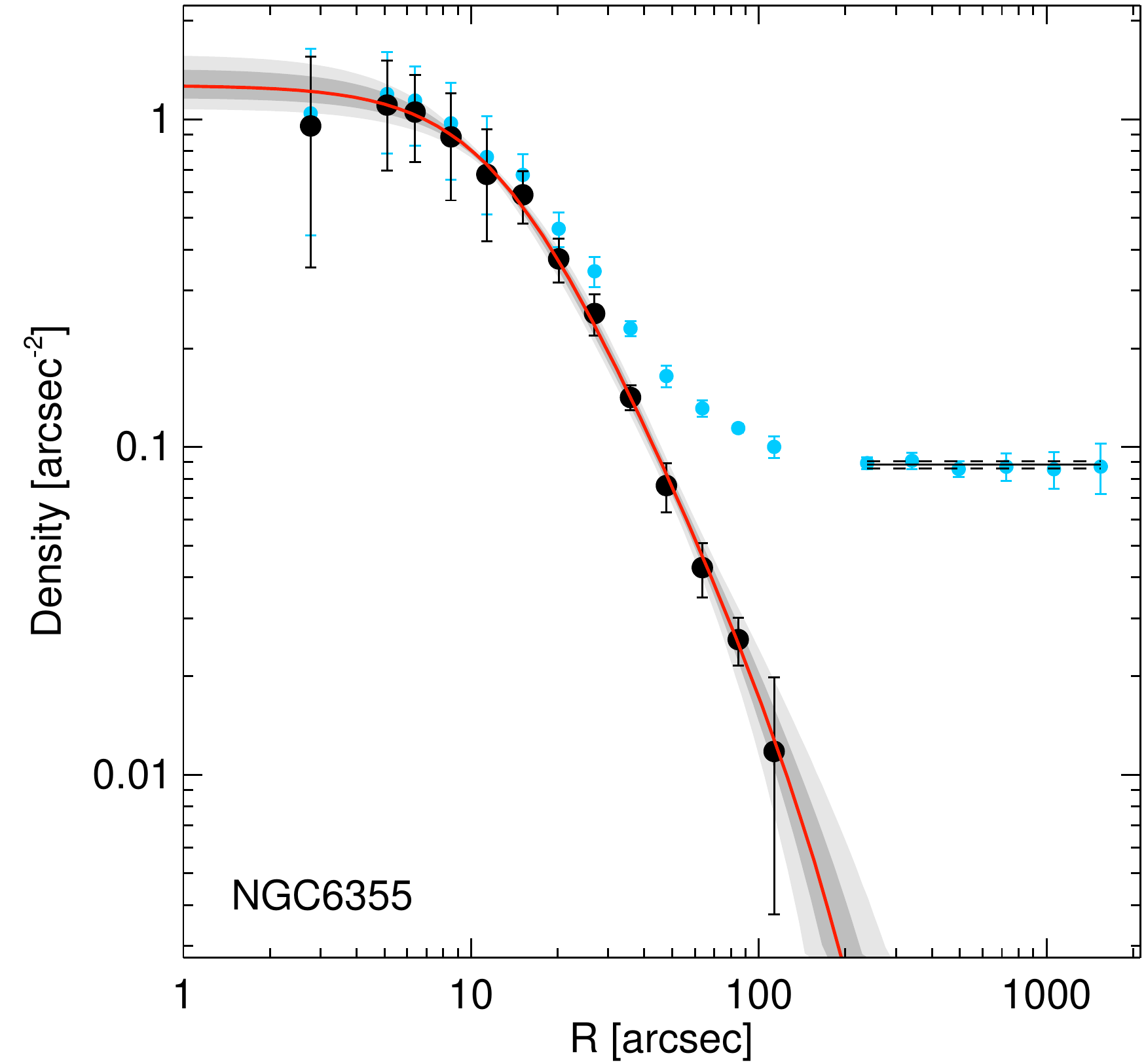}{0.33\textwidth}{}
          \fig{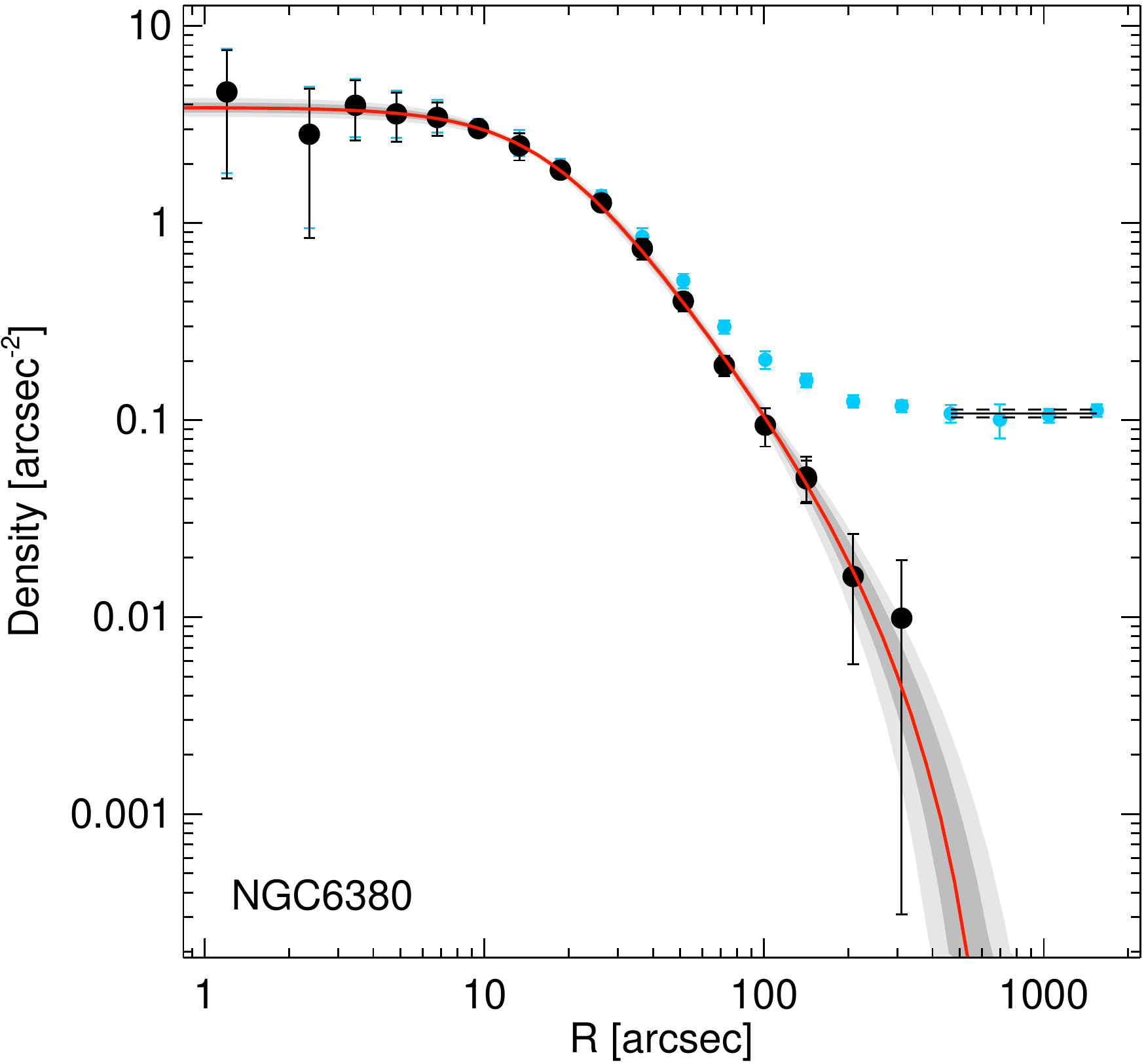}{0.33\textwidth}{}
          \fig{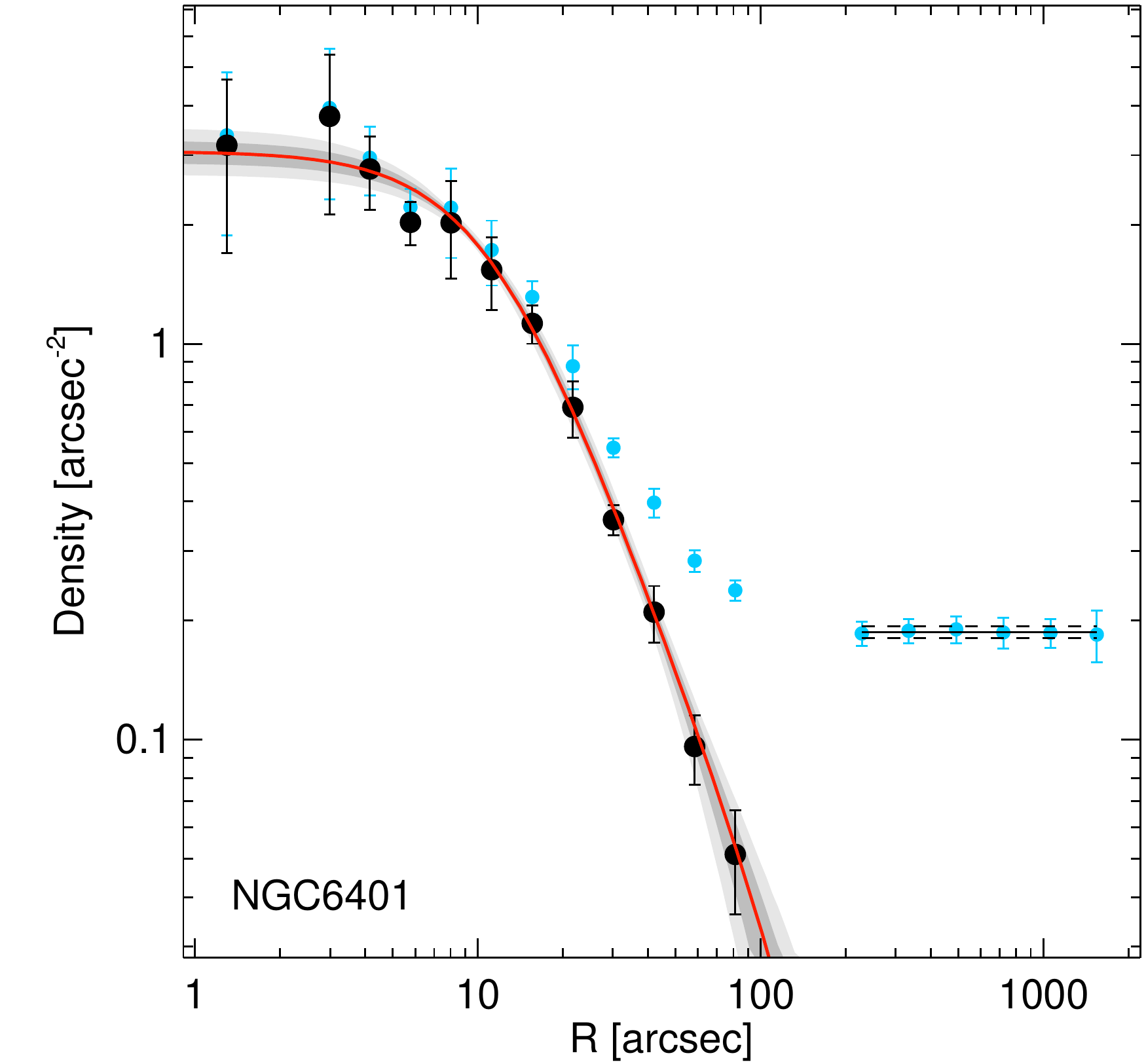}{0.33\textwidth}{}}
\gridline{\fig{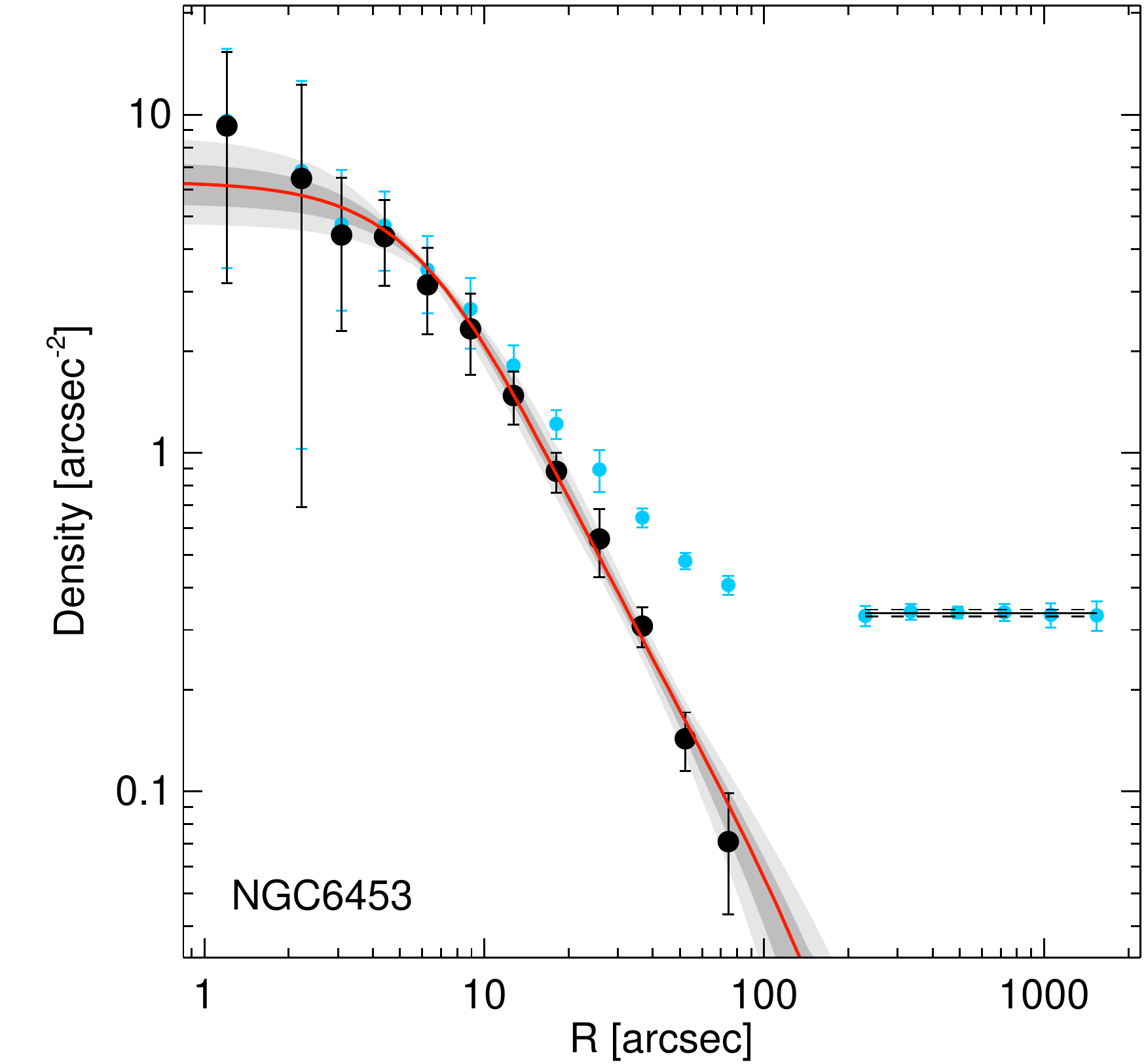}{0.33\textwidth}{}
	  \fig{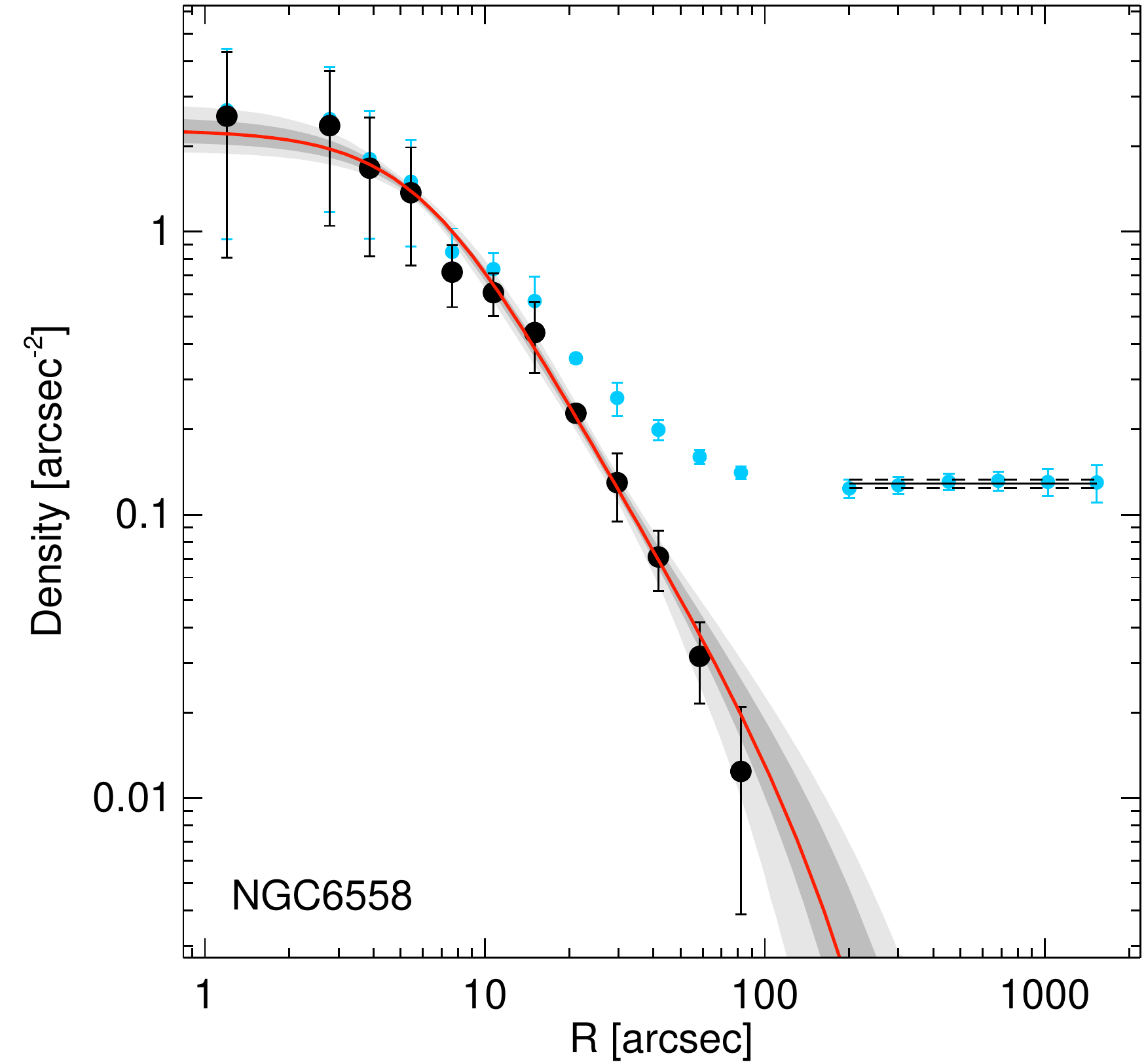}{0.33\textwidth}{}
	  \fig{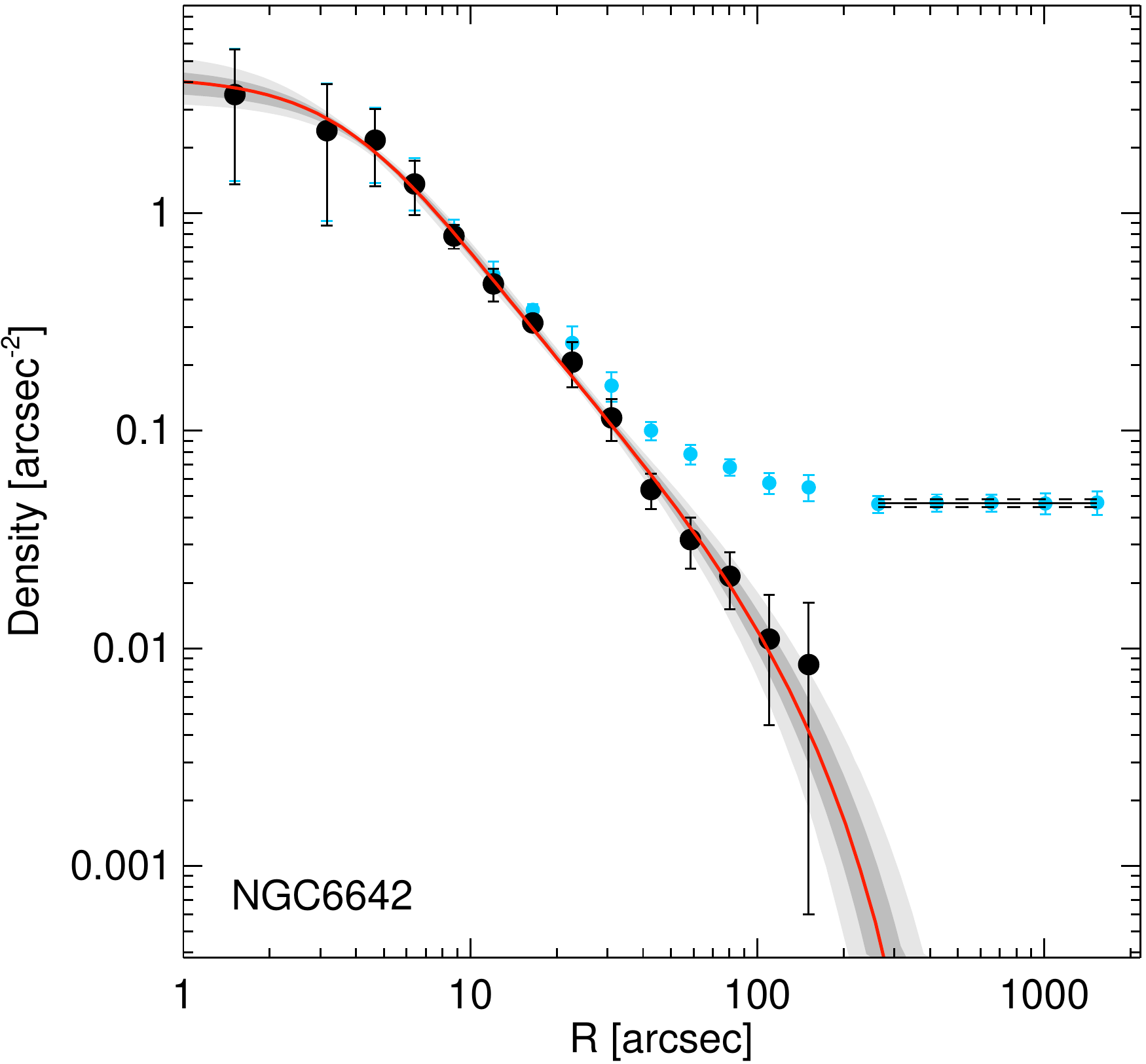}{0.33\textwidth}{}}
\caption{Radial density profiles for our target clusters.  Small cyan filled circles show densities before background subtraction, and the mean background and its uncertainty is shown using solid and dashed black horizontal lines respectively.  The best fit \citet{king66} profile is shown in red, and the dark and light grey shaded regions indicate the 1$\sigma$ and 2$\sigma$ uncertainties from the fit. 
\label{sbfig}}
\end{figure}

The best-fitting \citet{king66} profile for each of our target clusters is shown as a red line in Fig.~\ref{sbfig}, and the 1$\sigma$ (2$\sigma$) uncertainty range is illustrated using dark grey (light grey) shading.  While three of our target clusters (NGC 6355, 6380, 6401) are well-fit by the \citet{king66} profiles, the model fits underpredict the central density to some extent in nearly all of the other cases, and we return to this point in Sect.~\ref{densdiscusssect}.

\begin{deluxetable}{lcccccccccc}
\tablecaption{Cluster Centers and Results of King Profile Fits \label{sbtab}}
\tablehead{
\colhead{Cluster} & \colhead{RA (J2015.5)} & \colhead{Dec (J2015.5)} & \colhead{$\sigma_{pos}$} & \colhead{$W_{0}$} & \colhead{$r_{\rm 0}$} & \colhead{$r_{\rm t}$} & \colhead{$r_{\rm c,obs}$} & \colhead{$r_{\rm hl}$} & \colhead{$r_{\rm hm}$} & \colhead{$c$} \\ & deg & deg & $\arcsec$ & & $\arcsec$ & $\arcsec$ & $\arcsec$ & $\arcsec$ & $\arcsec$ &  
}
%%\colnumbers
\startdata
NGC6256 & 254.886107 & -37.120968 & 0.2 & $9.0^{+0.6}_{-0.6}$ & $10.3^{+1.0}_{-0.9}$ & $1317^{+499}_{-395}$ & $10.1^{+0.9}_{-0.9}$ & $109^{+51}_{-35}$ & $154^{+83}_{-55}$ & $2.12^{+0.15}_{-0.18}$ \\
NGC6325 & 259.496327 & -23.767677 & 0.6 & $9.2^{+0.4}_{-0.5}$ & $7.4^{+1.2}_{-1.2}$ & $1026^{+220}_{-189}$ & $7.3^{+1.2}_{-1.2}$ & $86^{+22}_{-17}$ & $124^{+35}_{-29}$ & $2.15^{+0.11}_{-0.12}$ \\
NGC6342 & 260.291573 & -19.587659 & 0.2 & $8.6^{+0.3}_{-0.3}$ & $6.0^{+0.6}_{-0.7}$ & $588^{+84}_{-79}$ & $5.8^{+0.6}_{-0.6}$ & $47^{+7}_{-6}$ & $64^{+12}_{-10}$ & $2.00^{+0.08}_{-0.10}$ \\
NGC6355 & 260.993533 & -26.352827 & 1.0 & $7.3^{+0.6}_{-0.6}$ & $13.7^{+1.5}_{-1.5}$ & $566^{+220}_{-150}$ & $13.1^{+1.3}_{-1.3}$ & $47^{+16}_{-10}$ & $62^{+20}_{-11}$ & $1.62^{+0.18}_{-0.17}$ \\
NGC6380 & 263.618611 & -39.069530 & 0.7 & $7.3^{+0.4}_{-0.4}$ & $18.8^{+1.5}_{-1.4}$ & $751^{+174}_{-138}$ & $18.0^{+1.3}_{-1.2}$ & $62^{+12}_{-9}$ & $83^{+15}_{-10}$ & $1.60^{+0.11}_{-0.11}$ \\
NGC6401 & 264.652191 & -23.909605 & 0.5 & $7.4^{+0.7}_{-0.7}$ & $12.2^{+1.2}_{-1.1}$ & $533^{+307}_{-169}$ & $11.7^{+1.0}_{-1.0}$ & $44^{+23}_{-11}$ & $58^{+30}_{-13}$ & $1.64^{+0.23}_{-0.20}$ \\
NGC6453 & 267.715508 & -34.598477 & 0.3 & $9.0^{+0.7}_{-0.8}$ & $7.4^{+1.3}_{-1.2}$ & $932^{+367}_{-310}$ & $7.2^{+1.2}_{-1.2}$ & $77^{+38}_{-27}$ & $109^{+62}_{-43}$ & $2.12^{+0.17}_{-0.23}$ \\
NGC6558 & 272.573974 & -31.764508 & 0.4 & $8.3^{+0.6}_{-0.6}$ & $7.1^{+0.8}_{-0.8}$ & $597^{+250}_{-188}$ & $6.9^{+0.7}_{-0.7}$ & $48^{+22}_{-14}$ & $64^{+35}_{-20}$ & $1.93^{+0.18}_{-0.19}$ \\
NGC6642 & 277.975957 & -23.475602 & 0.2 & $8.8^{+0.4}_{-0.4}$ & $4.5^{+0.6}_{-0.7}$ & $503^{+106}_{-91}$ & $4.4^{+0.6}_{-0.7}$ & $41^{+9}_{-7}$ & $57^{+15}_{-12}$ & $2.06^{+0.11}_{-0.10}$ \\
\enddata
%%%\tablecomments{Comments}
\end{deluxetable}

\section{Internal Kinematics \label{kinsect}}

\subsection{Quality Cuts and Sample Selection\label{cutsect}}

For the analysis of photometry and stellar density profiles in Sects.~\ref{censect}-\ref{denssect}, we applied a set of basic quality cuts to the photometry (listed in Sect.~\ref{obssect}) to remove sources which are non-stellar, spurious, and/or poorly characterized.  An analysis of the internal kinematics of our target clusters requires a much more stringent set of quality cuts, as blending and crowding can systematically bias the resulting kinematic profiles \citep[e.g.][]{bellini14,libralato_pm}.  Therefore we make the following additional quality cuts on the sample used for a kinematic analysis, in addition to those listed in Sect.~\ref{obssect}:

\begin{enumerate}[label=(\roman*)]
	\item We only retain stars with the best (highest) 50\% of QFIT values at their magnitude. \label{firstitem}
	\item We only retain stars that have measured proper motions based on $\geq$85\% of the exposures in which they were detected.
	\item We only retain stars that have a reduced $\chi^{2} \leq$ 1.3 in both coordinates from the proper motion fit.  
	\item We use a cut in relative proper motion to remove foreground and background disk and bulge stars.
	\item We retain only stars with proper motion uncertainties in each coordinate less than half of the local (measured using the nearest 75 stars in magnitude and distance from the cluster center) proper motion dispersion $\sigma_{\mu}$.
        \item We use a mild color cut around the cluster fiducial sequence to remove outliers and as an additional safeguard against any field contaminants that have proper motions similar to the cluster bulk motion. \label{lastitem}
\end{enumerate}

The impact of these cuts on the stellar sample is illustrated for an example case in Fig.~\ref{kinexamplefig}.  In the left panel, 
stars with measured proper motions are shown as green or black filled circles for field or cluster stars respectively, based on their relative proper motions shown in the vector point diagram in the upper right-hand panel of Fig.~\ref{kinexamplefig}.  In the lower right, we show a CMD zoomed in on the MSTO region, and also plot one-dimensional proper motion error versus magnitude.  Cluster members passing the basic cuts from Sect.~\ref{obssect} are again shown in black, as in the other panels, but now 
%showing cluster members passing the basic cuts from Sect.~\ref{obssect} in black.   Here, we plot 
stars that also pass the more stringent cuts listed in items \ref{firstitem}-\ref{lastitem} are shown in blue.  Lastly, we place bright and faint magnitude limits on our high-quality sample, shown in red in Fig.~\ref{kinexamplefig}.  Specifically, we require F814W magnitudes brighter than 0.5 mag faintward of the MSTO, which has the advantage of providing, again, a mono-mass sample for kinematic analysis consistently across our target clusters.  Meanwhile, we impose a bright limit by avoiding the magnitude range where stars are detected in only short exposures, which inflates their proper motion uncertainties \citep[e.g.][]{raso1261}.  While Fig.~\ref{kinexamplefig} represents just one example out of our nine target clusters, it is a typical case, and it bears mention that in \textit{all} cases, the archival first epoch imaging is the limiting factor in terms of not only photometric depth 
but also astrometric precision: The first epoch imaging is undithered and consists of fewer individual exposures than our second epoch imaging (this is why we require fairly stringent cuts on the reduced $\chi^{2}$ values from the proper motion fits).    

\subsection{Proper Motion Dispersion \label{vdispsect}}

We measure the proper motion dispersion profiles of our target clusters using our high-quality mono-mass sample (shown in red in the lower right panel of Fig.~\ref{kinexamplefig} for an example case), and divide this sample into radial bins that serve as a compromise between preserving spatial resolution while remaining adequately populated following \citet{watkinskin}.  Within each radial bin, the proper motion dispersion $\sigma_{\mu}$ is calculated by assuming that the proper motion follows a Gaussian distribution centered around a cluster mean value with a total observed scatter resulting from the quadrature sum of the observational uncertainties on individual proper motions plus the proper motion dispersion \citep[e.g.][]{walker06,vandermarel10,watkinskin,raso1261}.  The only two free parameters are the mean proper motion and the proper motion dispersion, and we solve for these two parameters using a maximum likelihood approach with \texttt{emcee}, quoting the median of the posterior distribution as the best-fit value and the 16th and 84th percentiles as the uncertainties.  This calculation includes the correction of \citet{vandeven} to account for the underestimate of the dispersion using maximum likelihood estimators, although this correction is essentially insignificant and affects our measured $\sigma_{\mu}$ values by $<$0.5\% in the majority of cases (and $<$3\% in all cases), well within their uncertainties.

\begin{figure}
\gridline{\fig{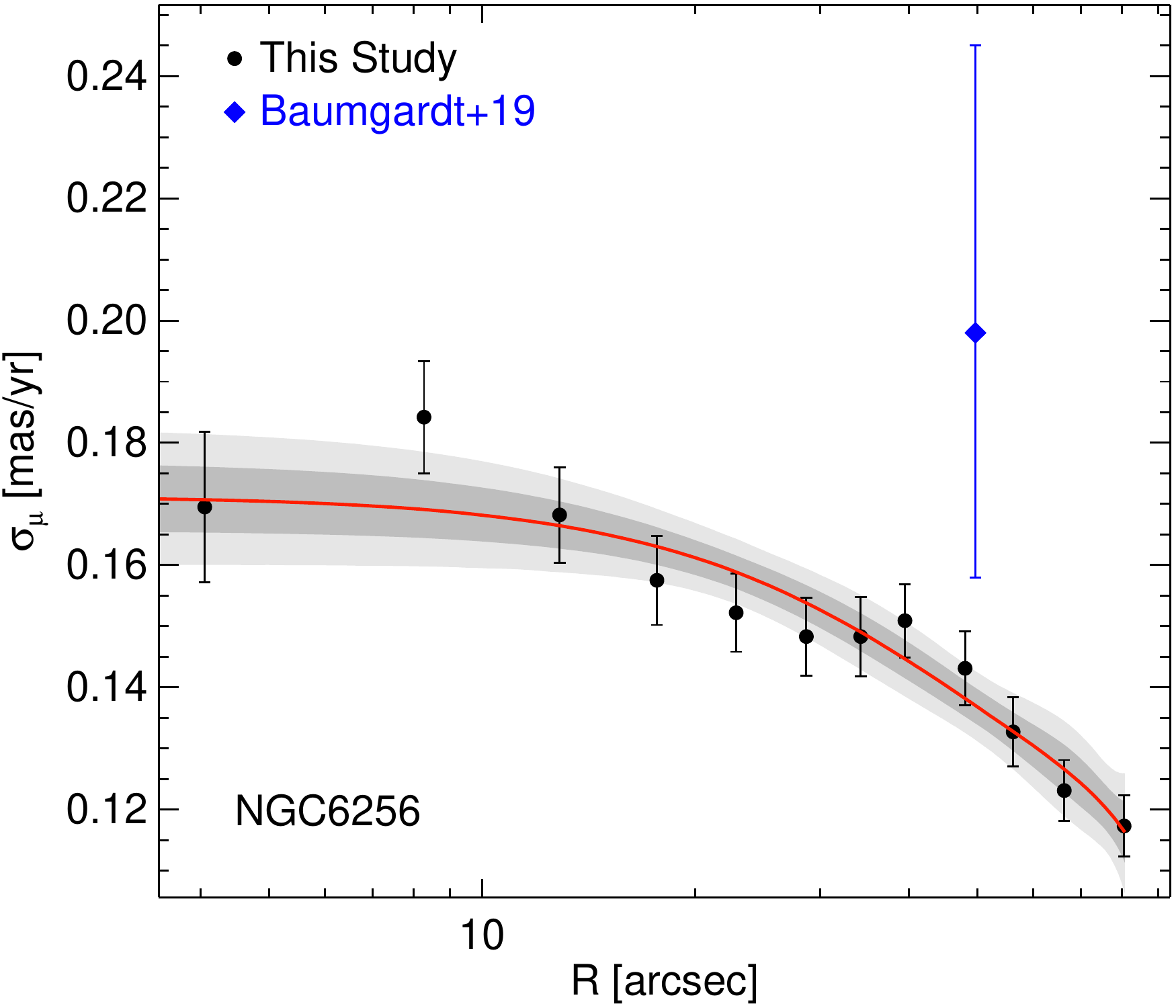}{0.33\textwidth}{}
	  \fig{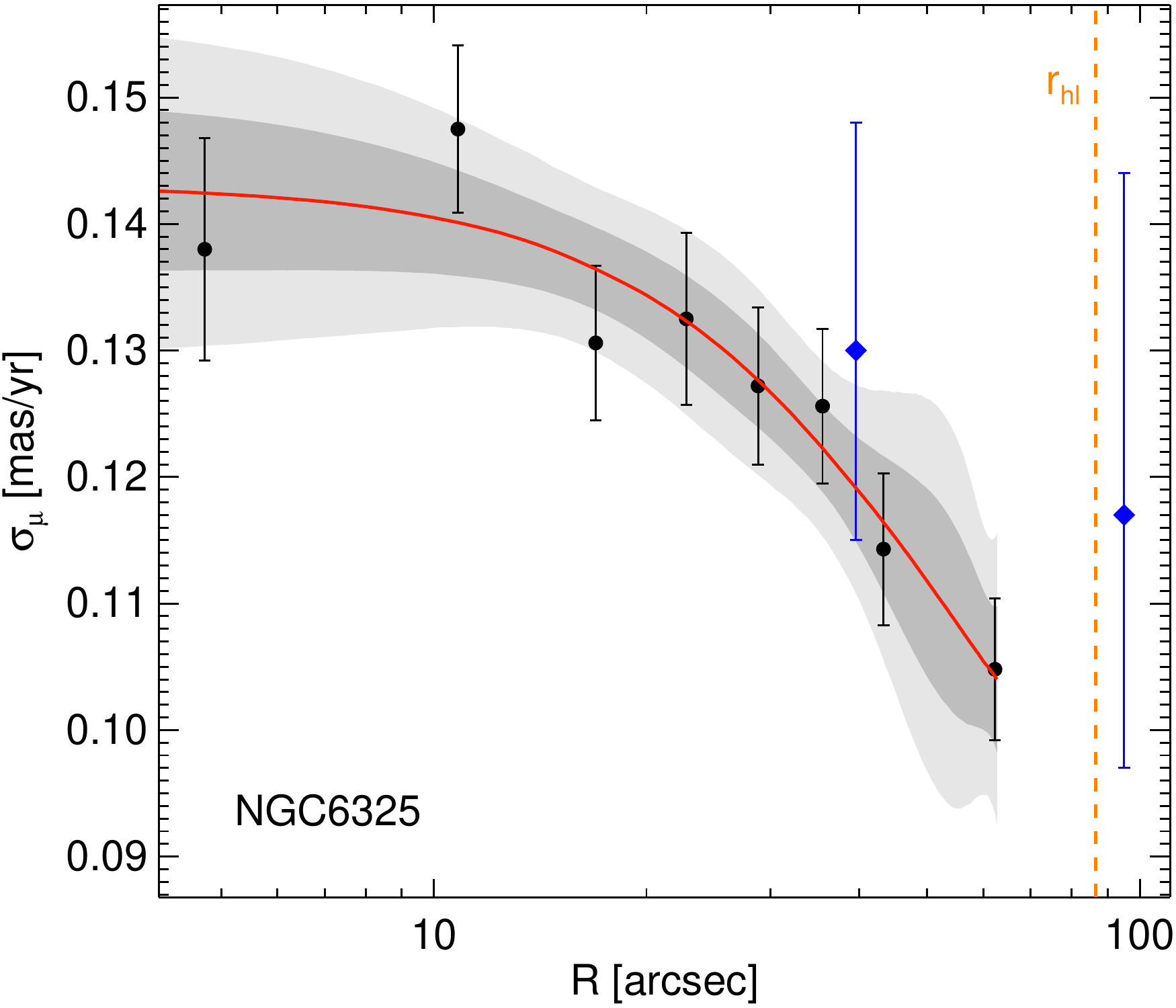}{0.33\textwidth}{}
          \fig{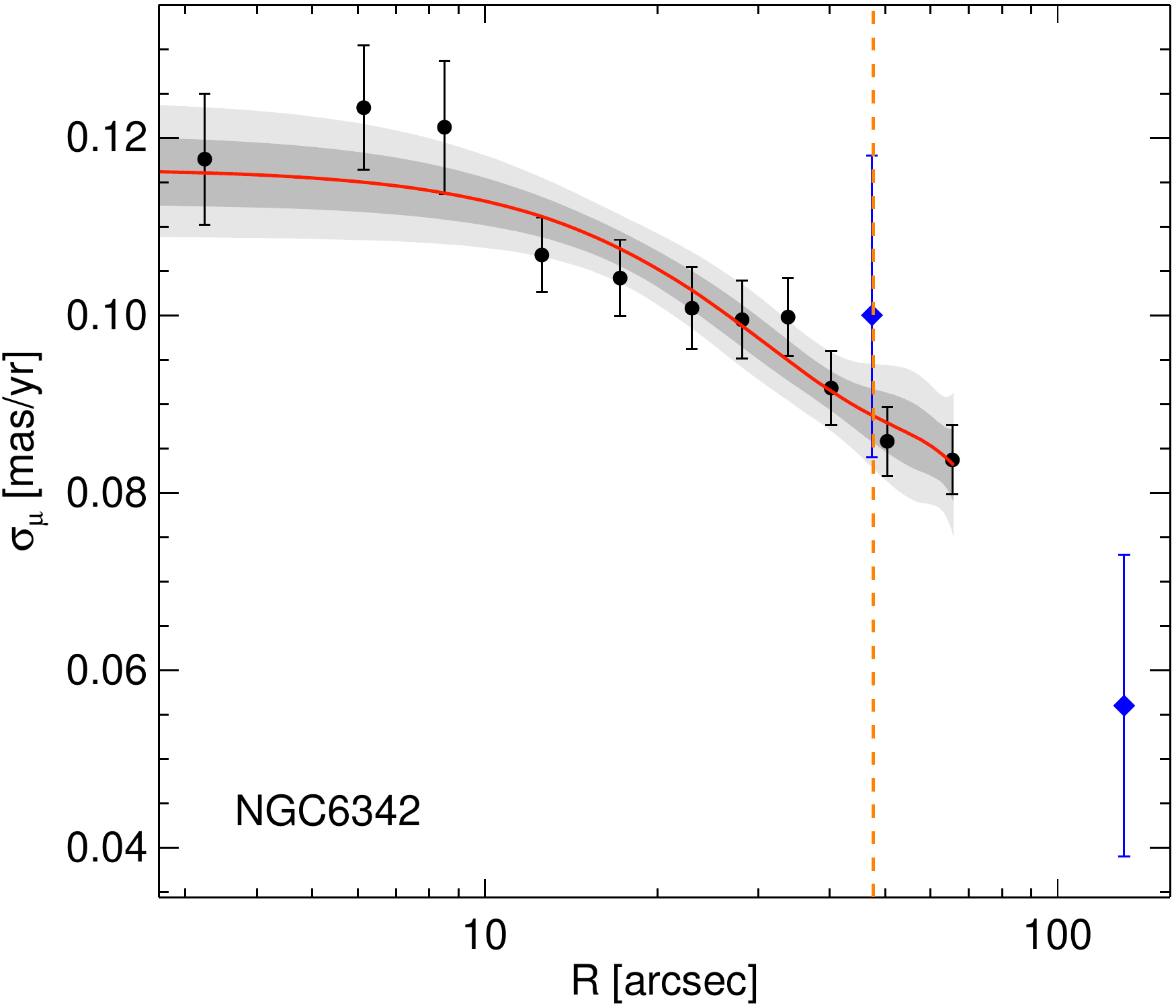}{0.33\textwidth}{}}
\gridline{\fig{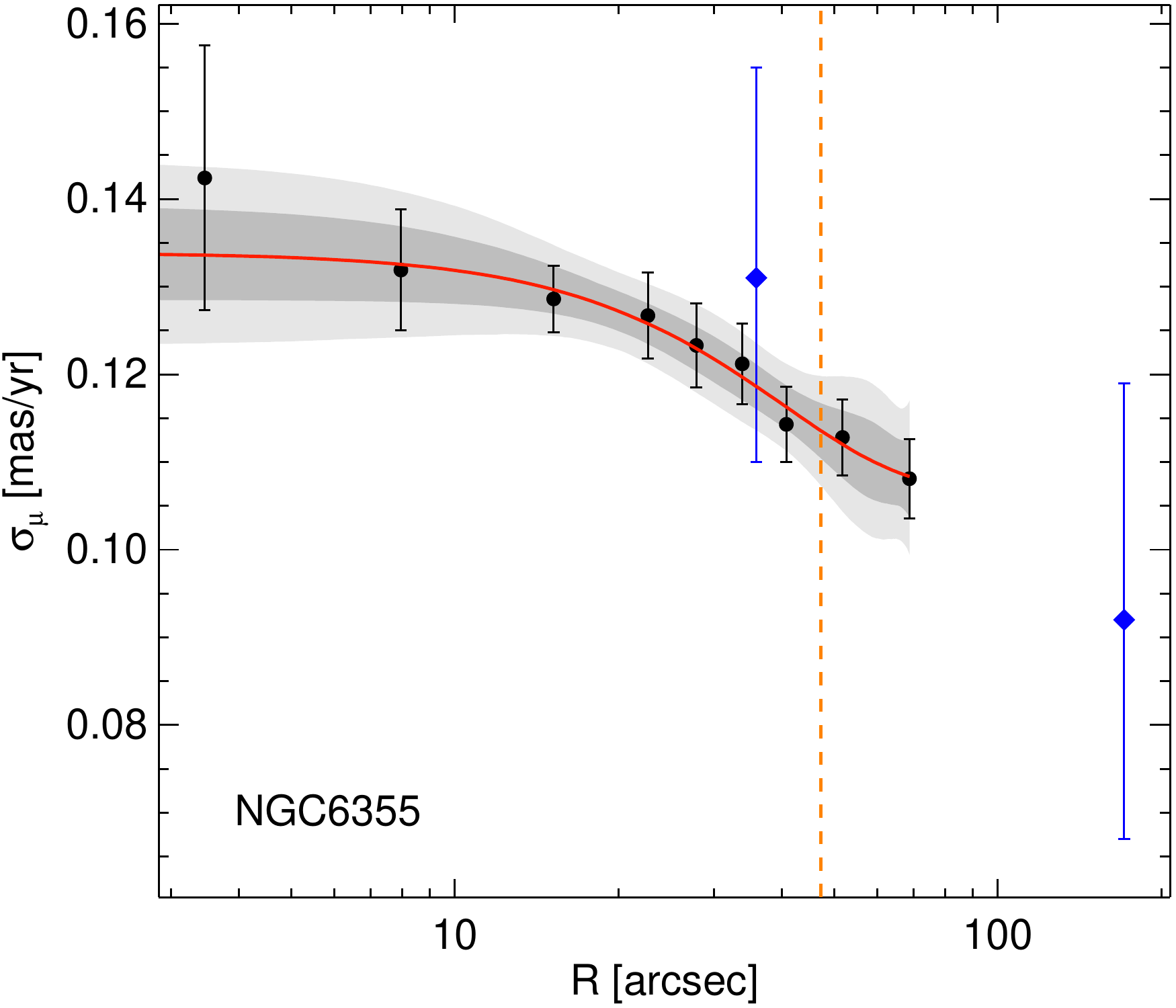}{0.33\textwidth}{}
          \fig{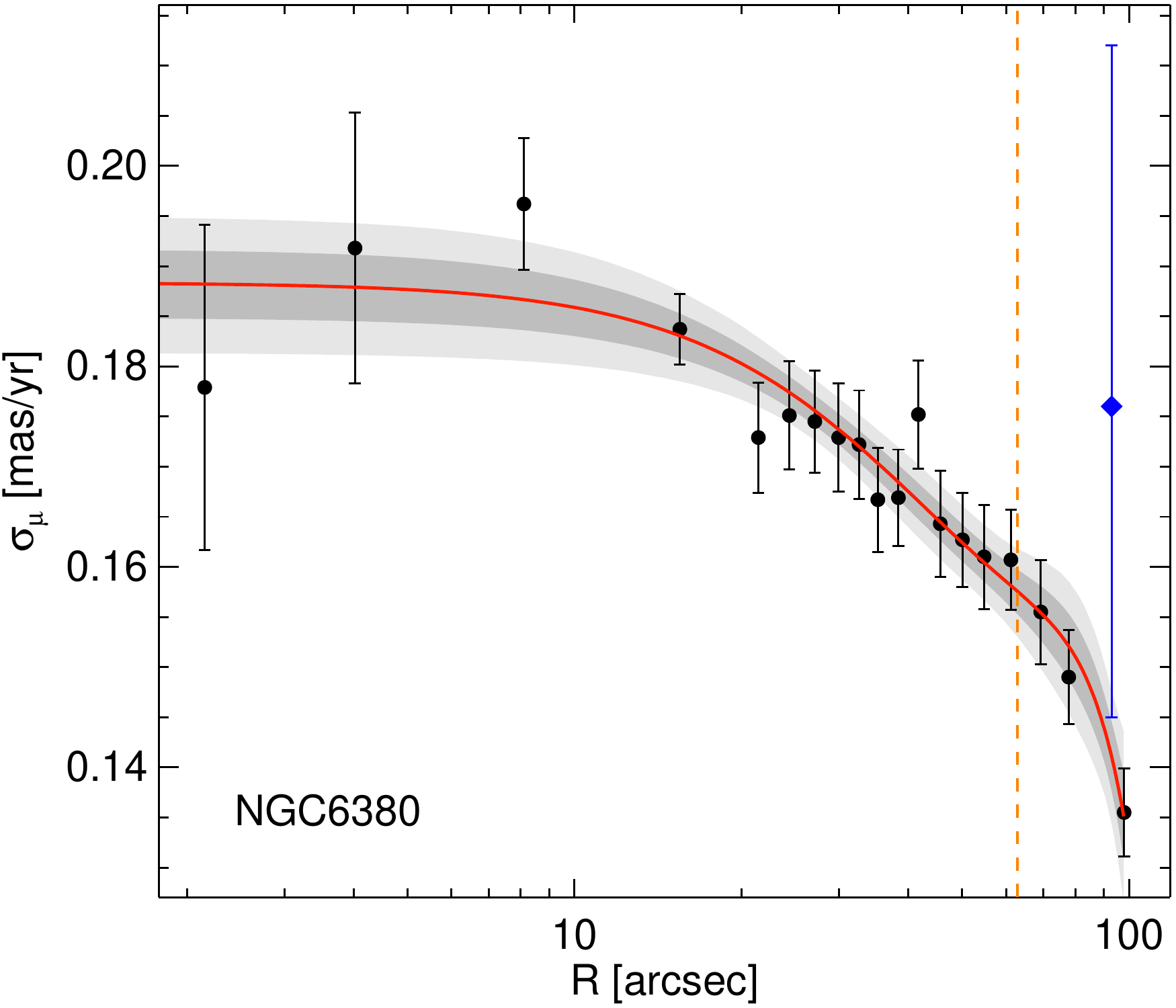}{0.33\textwidth}{}
          \fig{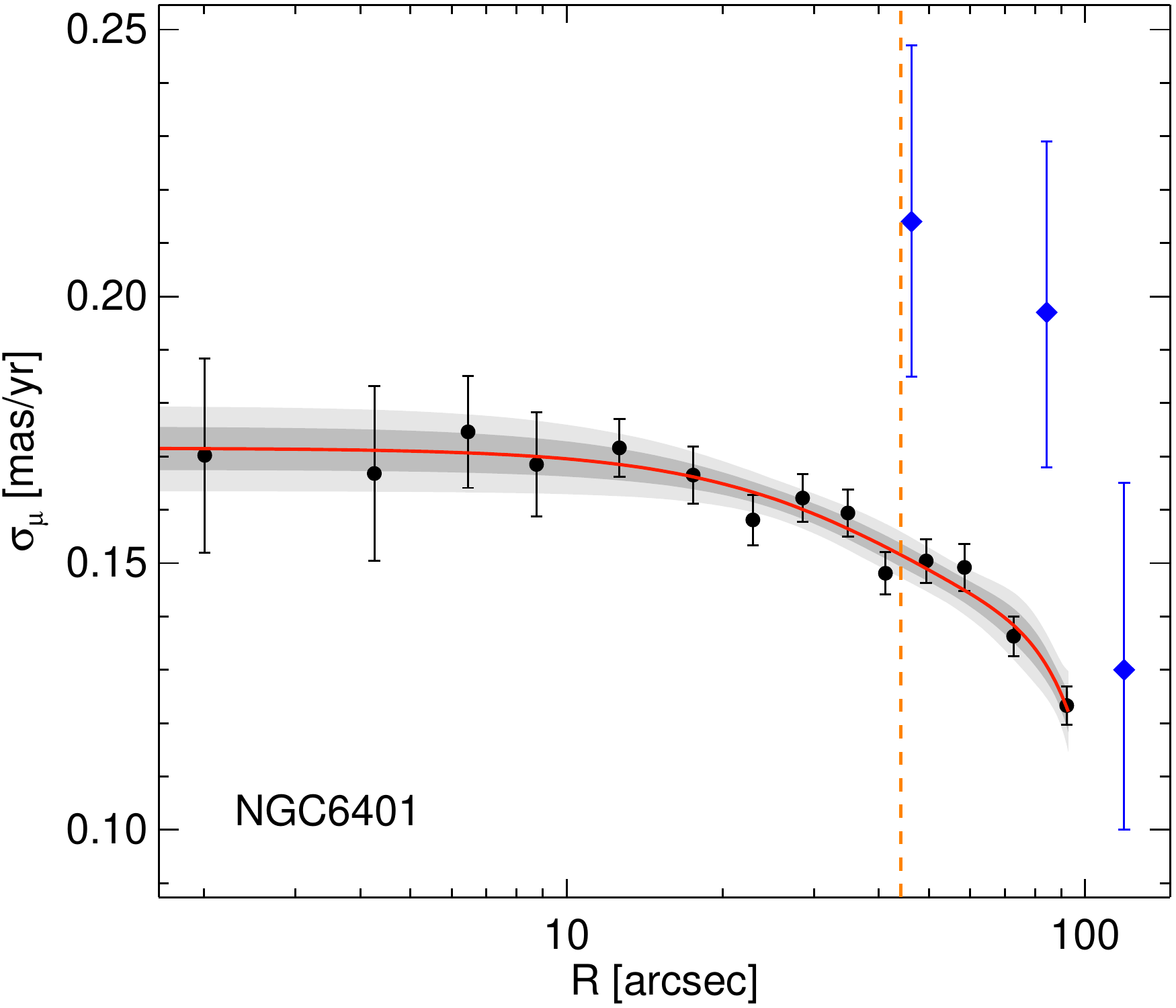}{0.33\textwidth}{}}
\gridline{\fig{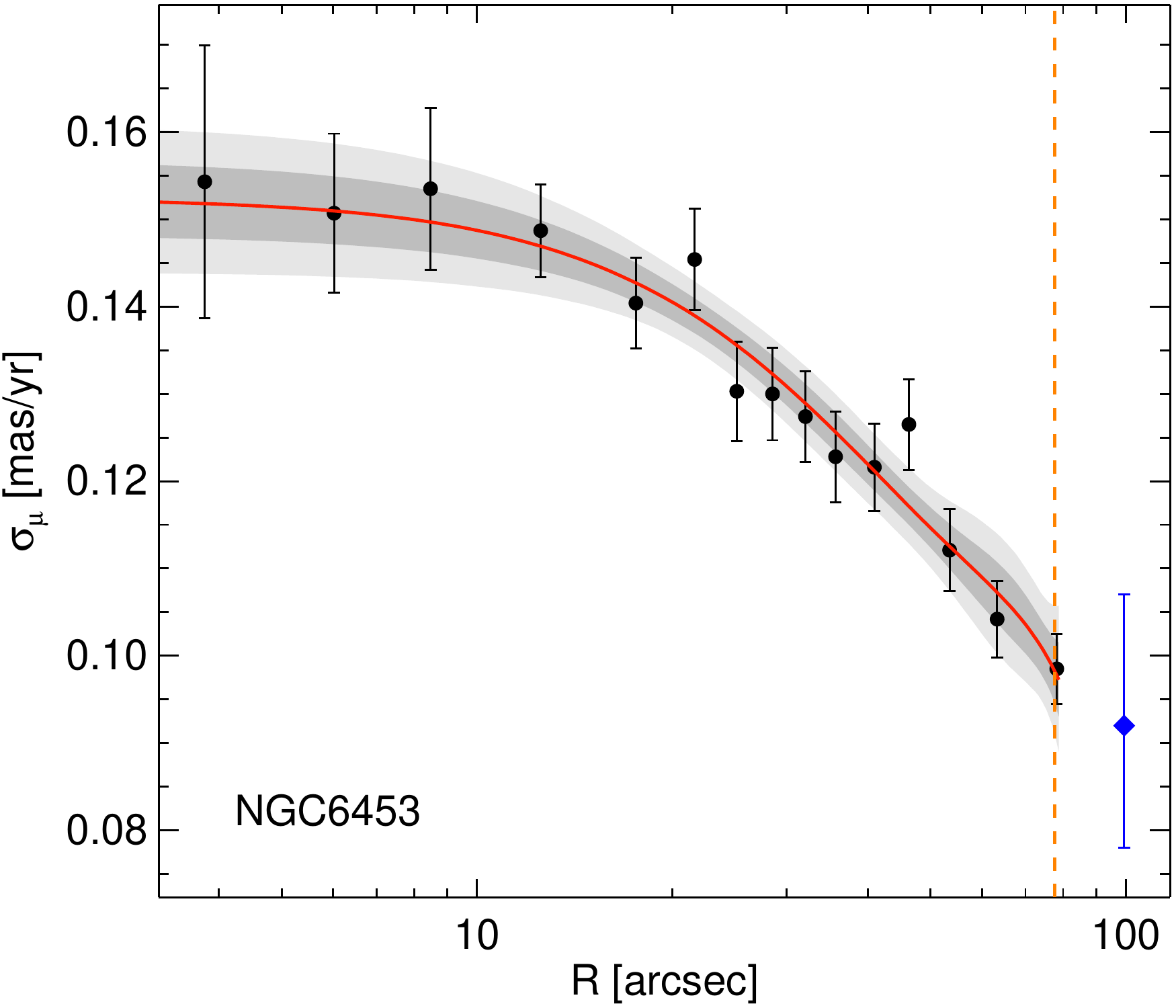}{0.33\textwidth}{}
	  \fig{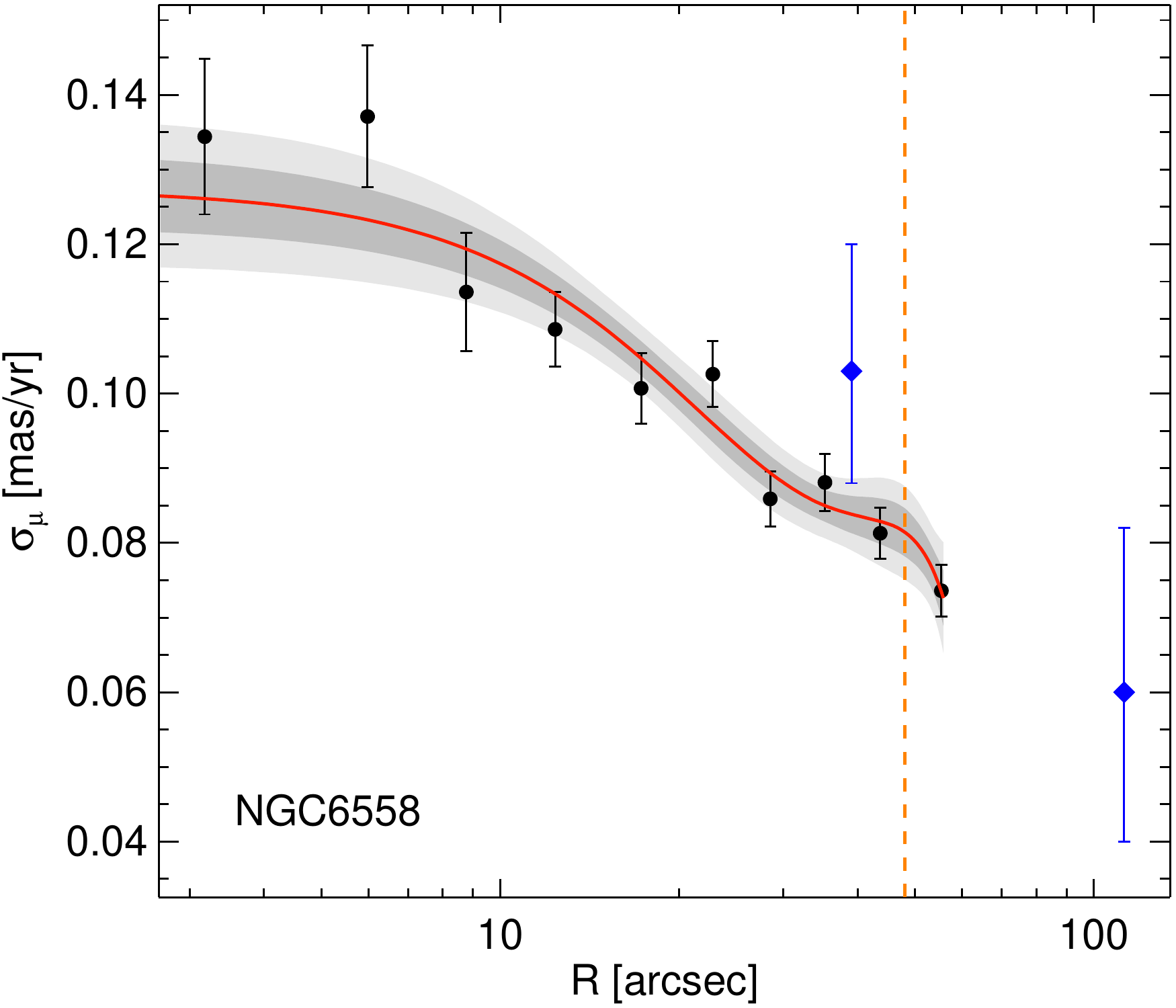}{0.33\textwidth}{}
	  \fig{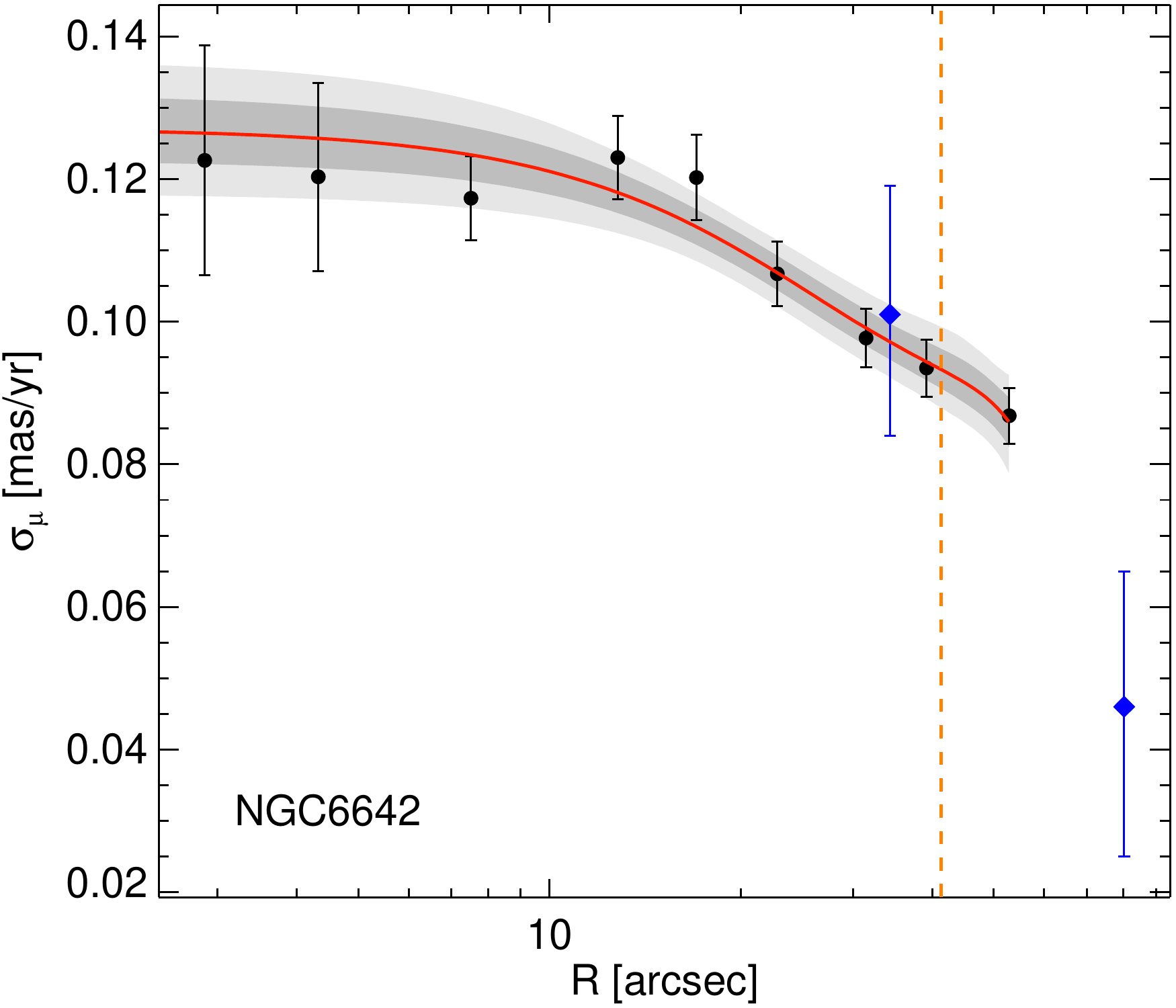}{0.33\textwidth}{}}
\caption{Proper motion dispersion profiles versus radius for our target clusters.  Fourth order monotonic, decreasing polynomial fits, constrained to be flat at the cluster centers, are shown in red, and their 1-$\sigma$ (2-$\sigma$) uncertainties are shown using dark (light) grey shading.
Additional data from the \textit{Gaia} DR2-based proper motion dispersion profiles of \citet{b19} are shown in blue, and the orange vertical dashed lines indicate the value of $r_{\rm hl}$ for each cluster from our density profile fits in Sect.~\ref{denssect}.  
\label{vdisprfig}}
\end{figure}

We then fit a 4th order monotonic decreasing polynomial, constrained to be flat at zero radius, to proper motion dispersion as a function of radius following \citet{watkinskin}, also using \texttt{emcee}.  These fits are shown, along with their uncertainties, in Fig.~\ref{vdisprfig}.  For comparison, we also overplot \textit{Gaia} DR2-based proper motion dispersions from \citet{b19} in blue.  While we caution against blindly extrapolating our fits beyond their range of validity, they generally compare well with the proper motion dispersions in \citet{b19}.  The minority of cases where the \citet{b19} values are discrepant with our fits beyond their (necessarily larger) uncertainties (most notably NGC 6401, and to a lesser extent NGC 6256 and NGC 6380) correspond to the more crowded, extincted sightlines within our sample, making it more difficult to disentangle cluster and field stars using \textit{Gaia} DR2, which would cause the resultant velocity dispersions to be overestimated, consistent with Fig.~\ref{vdisprfig}.  The results of our polynomial fits are given in Table \ref{vdisptab}, where the fit values and their uncertainties from our maximum likelihood fits are given for the cluster center, as well as at the King radius $r_{\rm 0}$\footnote{We use $r_{\rm 0}$ here rather than the half-power radius $r_{\rm c,obs}$ to allow a direct comparison with \citet{watkinskin}.}, 0.5 times the half-light radius, and the half-light radius $r_{\rm hl}$ from our density profile fits in Sect.~\ref{denssect}.  Our velocity dispersion (and anisotropy) profiles are available electronically via Table \ref{vadatatab}.   

\begin{deluxetable}{lcccc}
\tablecaption{Proper Motion Dispersion at Selected Clustercentric Radii \label{vdisptab}}
\tablehead{
\colhead{Cluster} & \colhead{$\sigma_{\mu}(r=0)$} & \colhead{$\sigma_{\mu}(r_{\rm 0})$} & \colhead{$\sigma_{\mu}(0.5r_{\rm hl})$} & \colhead{$\sigma_{\mu}(r_{\rm hl})$}  \\ & mas yr$^{-1}$ & mas yr$^{-1}$ & mas yr$^{-1}$ & mas yr$^{-1}$
}
%%\colnumbers
\startdata
NGC6256 & $0.171^{+0.006}_{-0.006}$ & $0.168^{+0.004}_{-0.004}$ & $0.134^{+0.003}_{-0.003}$ &   \\
NGC6325 & $0.143^{+0.007}_{-0.007}$ & $0.142^{+0.005}_{-0.005}$ & $0.116^{+0.005}_{-0.006}$ &   \\
NGC6342 & $0.116^{+0.004}_{-0.004}$ & $0.115^{+0.003}_{-0.003}$ & $0.102^{+0.002}_{-0.002}$ & $0.089^{+0.003}_{-0.003}$ \\
NGC6355 & $0.134^{+0.005}_{-0.005}$ & $0.130^{+0.003}_{-0.003}$ & $0.125^{+0.002}_{-0.002}$ & $0.114^{+0.003}_{-0.003}$ \\
NGC6380 & $0.188^{+0.003}_{-0.004}$ & $0.181^{+0.002}_{-0.002}$ & $0.173^{+0.002}_{-0.002}$ & $0.158^{+0.002}_{-0.002}$ \\
NGC6401 & $0.172^{+0.004}_{-0.004}$ & $0.169^{+0.003}_{-0.003}$ & $0.164^{+0.002}_{-0.002}$ & $0.152^{+0.002}_{-0.002}$ \\
NGC6453 & $0.152^{+0.004}_{-0.004}$ & $0.150^{+0.004}_{-0.004}$ & $0.123^{+0.002}_{-0.002}$ & $0.098^{+0.004}_{-0.004}$ \\
NGC6558 & $0.127^{+0.005}_{-0.005}$ & $0.122^{+0.004}_{-0.004}$ & $0.094^{+0.002}_{-0.003}$ & $0.081^{+0.003}_{-0.003}$ \\
NGC6642 & $0.127^{+0.005}_{-0.004}$ & $0.126^{+0.004}_{-0.004}$ & $0.109^{+0.002}_{-0.002}$ & $0.093^{+0.003}_{-0.003}$ \\
\enddata
\tablecomments{The values given here are determined directly from maximum likelihood fits of a 4th order decreasing polynomial (constrained to be flat in the center) to the observed proper motion dispersion profiles as described in Sect.~\ref{vdispsect}.} 
\end{deluxetable}

\begin{deluxetable}{lccccccccc}
  \tablecaption{Radial Proper Motion Dispersion and Anisotropy Profiles \label{vadatatab}}
  \tablehead{
    \colhead{Cluster} & \colhead{Bin} & \colhead{$r_{min}$} & \colhead{$r$} & \colhead{$r_{max}$} & \colhead{N} & \colhead{$\sigma_{\mu}$} & \colhead{$\sigma_{\rm tan}/\sigma_{\rm rad}$} \\ & & $\arcsec$ & $\arcsec$ & $\arcsec$ & & mas yr$^{-1}$ & 
  }
\startdata
NGC6256 & 1 &   0.646 &   4.055 &   5.960 &  49 &  0.170$\pm$0.012 &  0.856$\pm$0.126 \\
NGC6256 & 2 &   6.064 &   8.273 &   9.988 &  97 &  0.184$\pm$0.009 &  0.951$\pm$0.104 \\
NGC6256 & 3 &  10.000 &  12.857 &  15.216 & 128 &  0.168$\pm$0.008 &  1.033$\pm$0.095 \\
NGC6256 & 4 &  15.238 &  17.645 &  19.948 & 128 &  0.158$\pm$0.007 &  0.994$\pm$0.085 \\
NGC6256 & 5 &  20.144 &  22.831 &  25.563 & 153 &  0.152$\pm$0.006 &  0.947$\pm$0.083 \\
\enddata
\tablecomments{Table \ref{vadatatab} is published in its entirety in machine-readable format.  A portion is shown here for guidance regarding its form and content.}
\end{deluxetable}

\subsection{Anisotropy \label{anisosect}}

\begin{figure}
\gridline{\fig{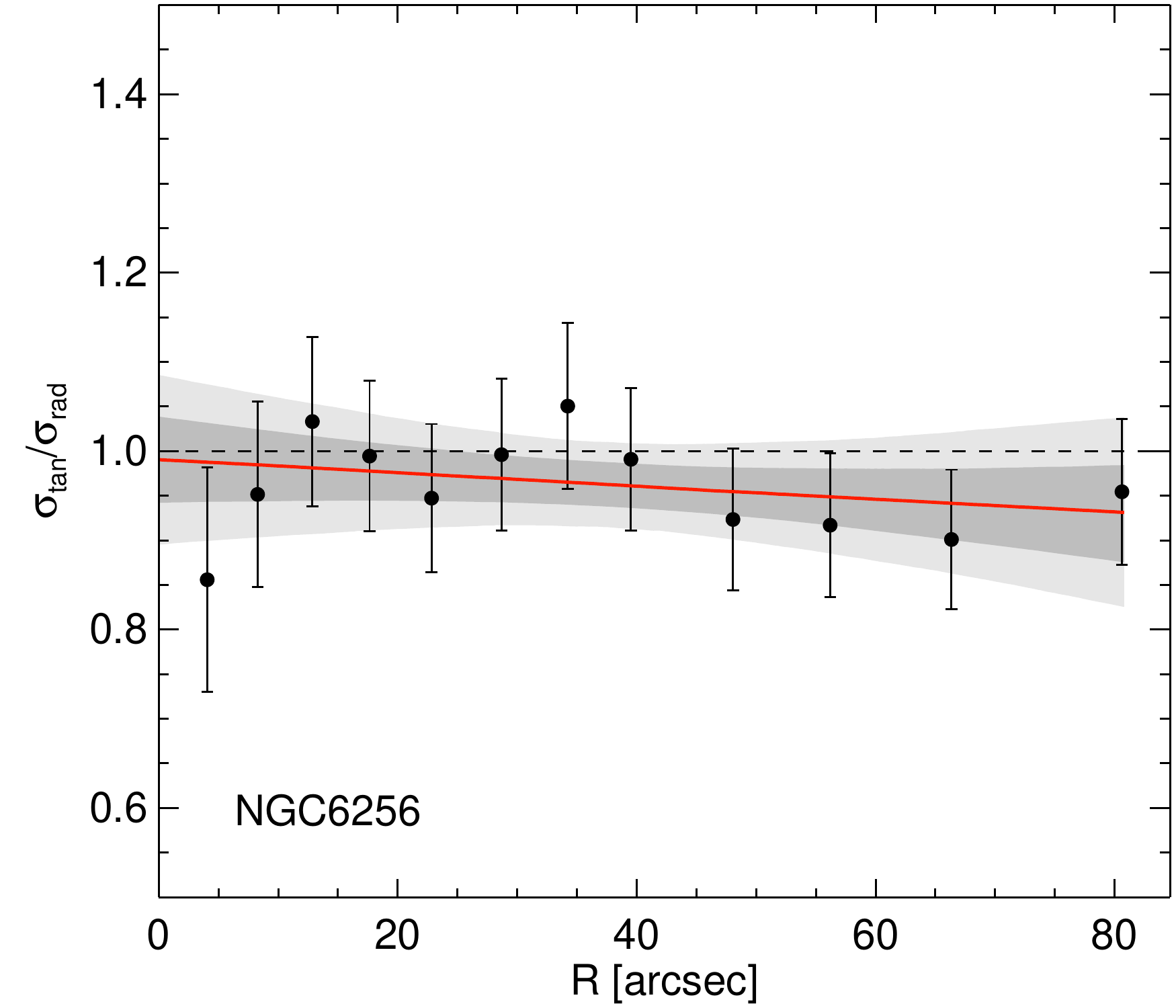}{0.33\textwidth}{}
	  \fig{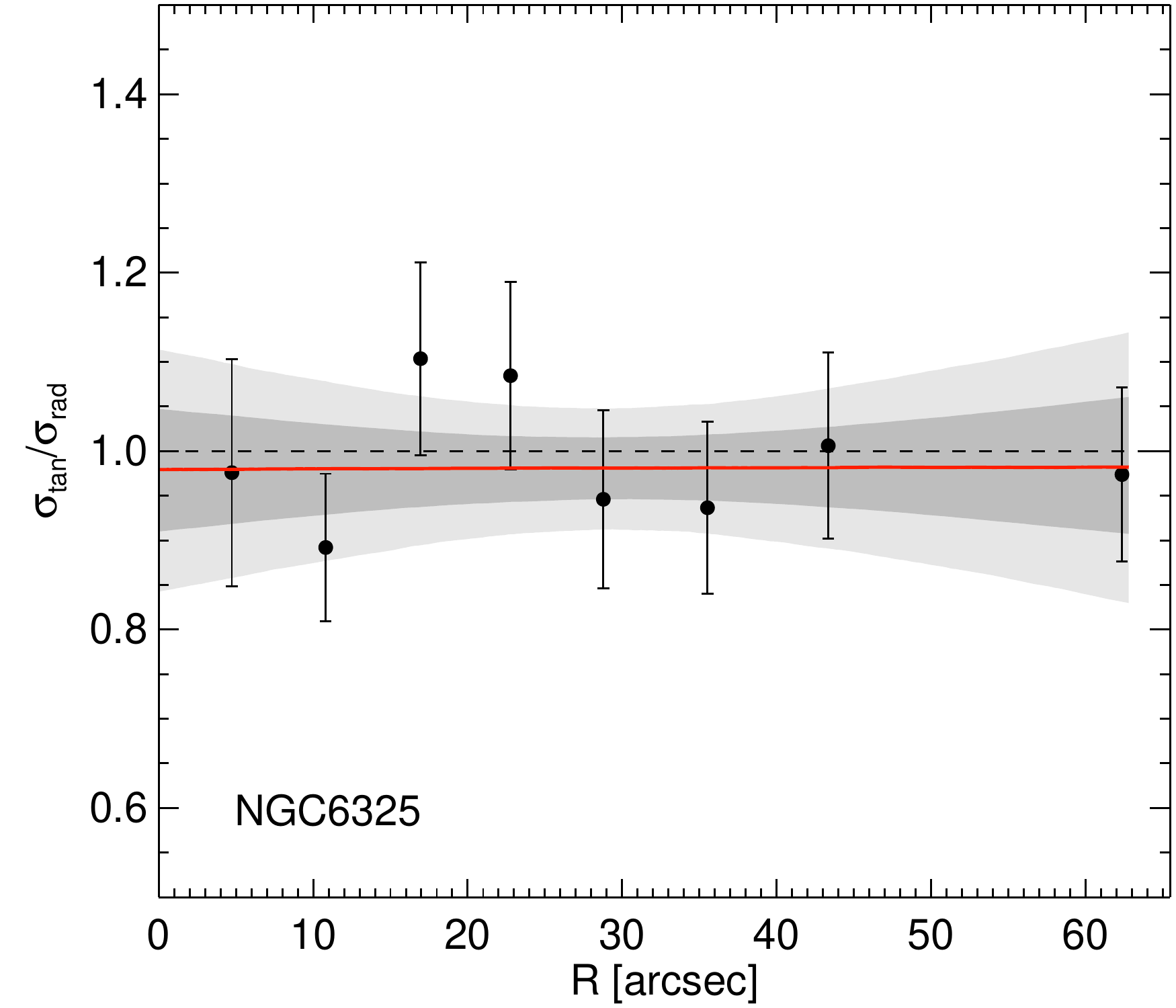}{0.33\textwidth}{}
          \fig{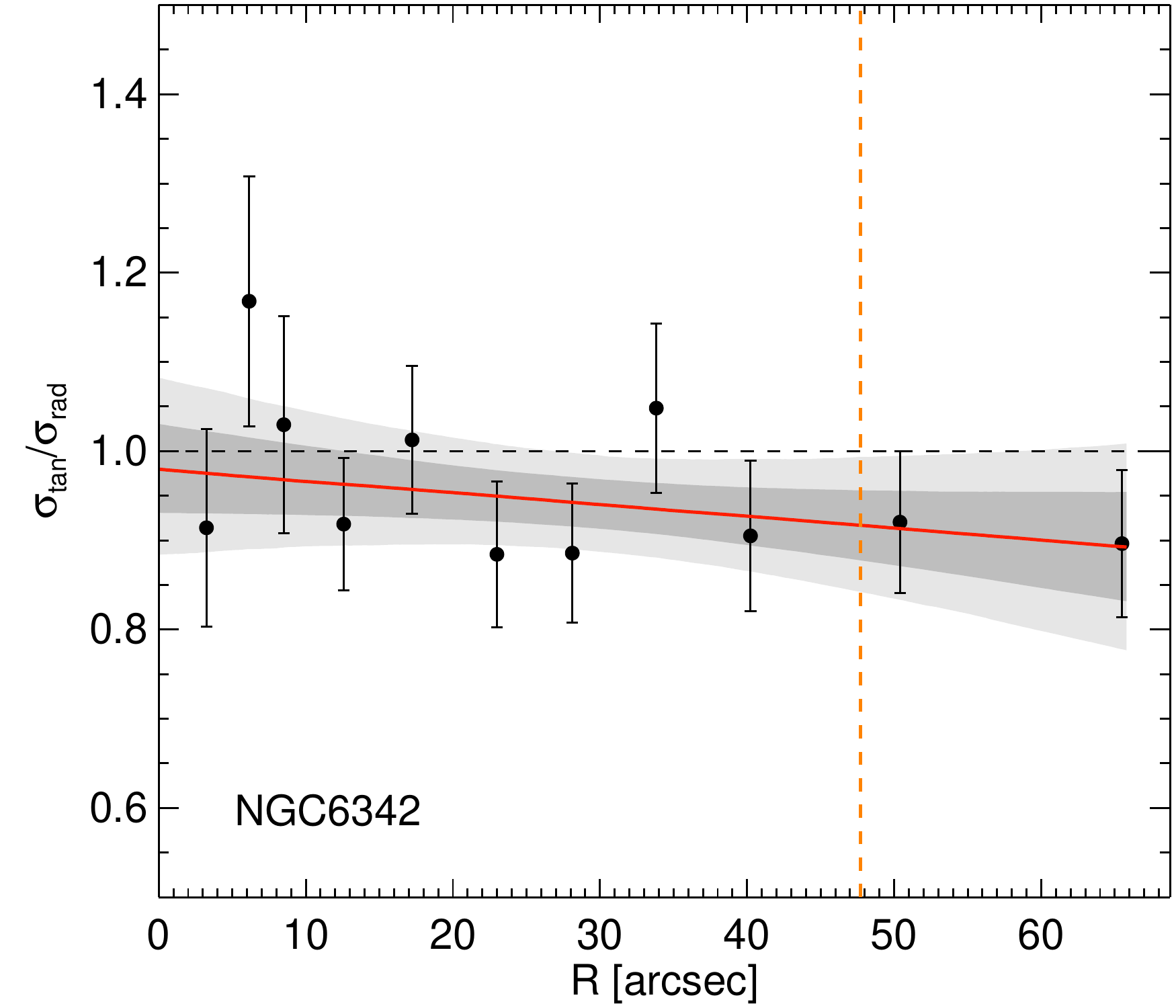}{0.33\textwidth}{}}
\gridline{\fig{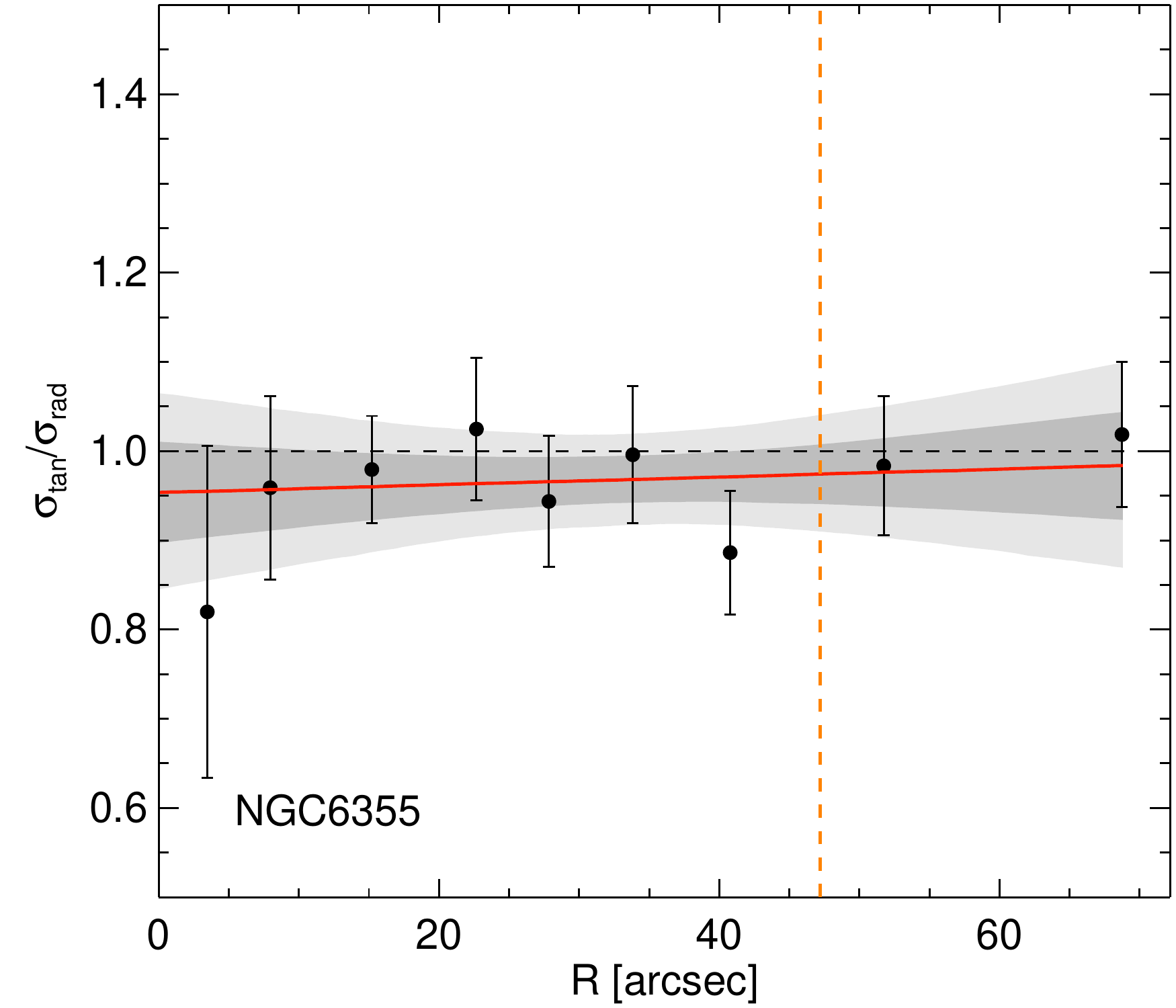}{0.33\textwidth}{}
          \fig{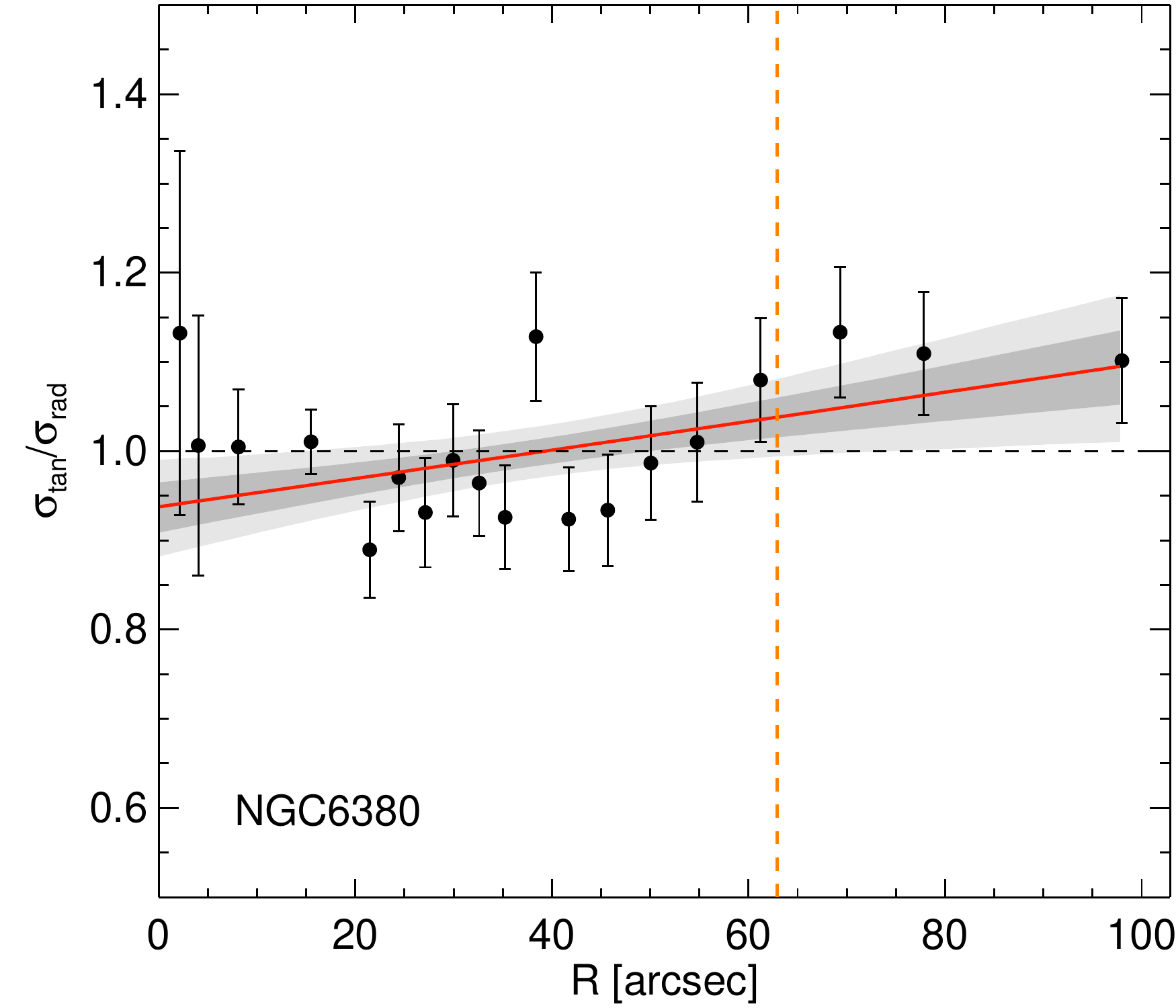}{0.33\textwidth}{}
          \fig{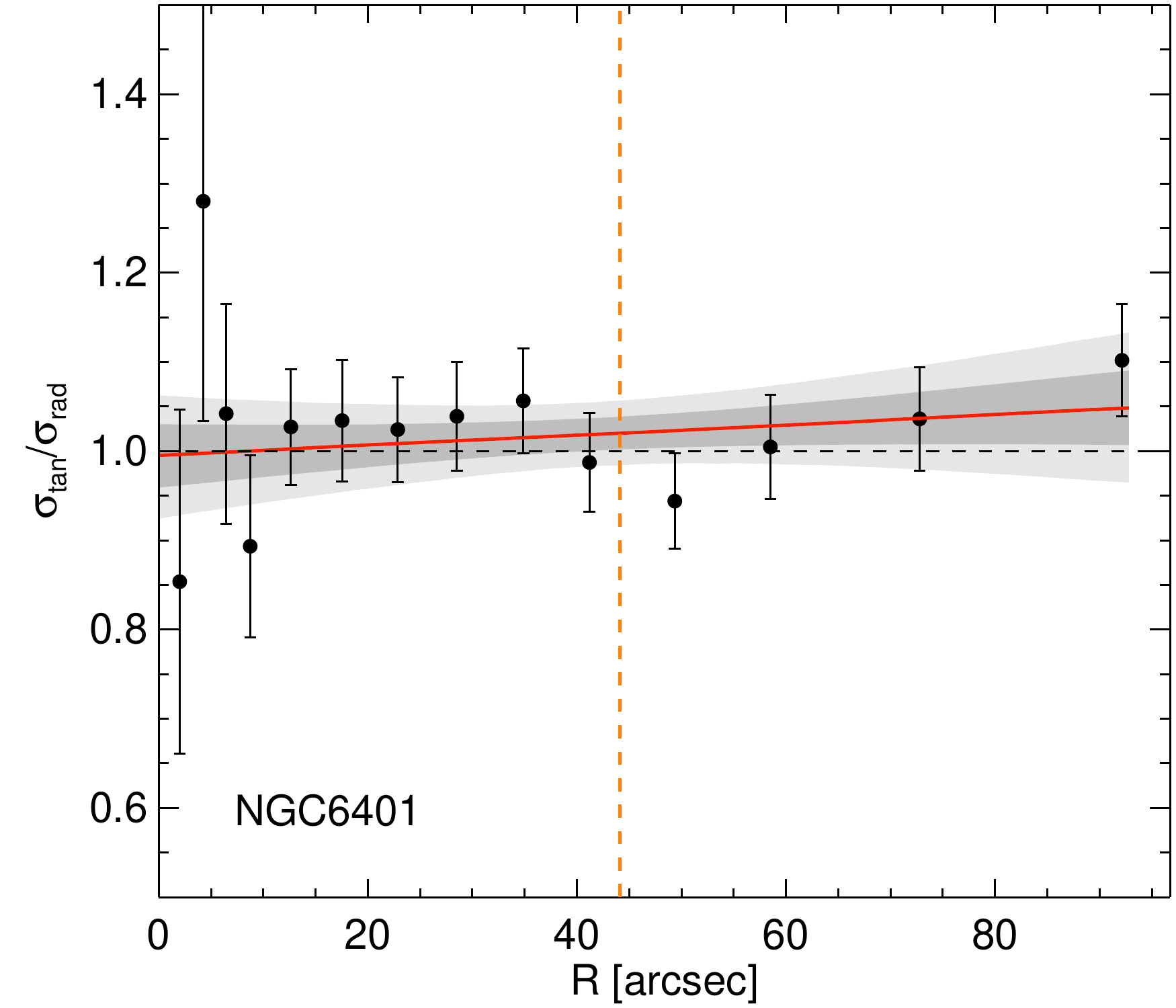}{0.33\textwidth}{}}
\gridline{\fig{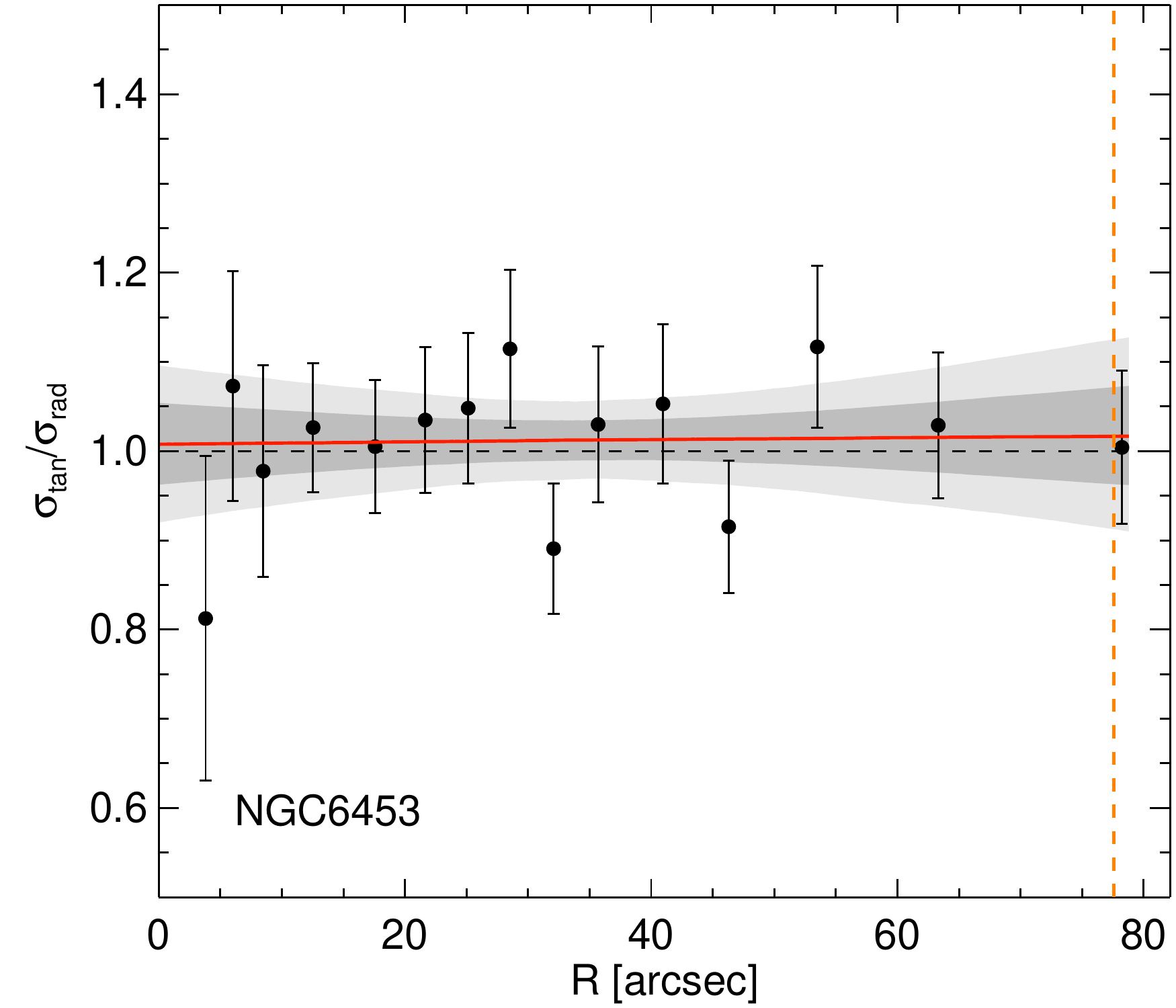}{0.33\textwidth}{}
	  \fig{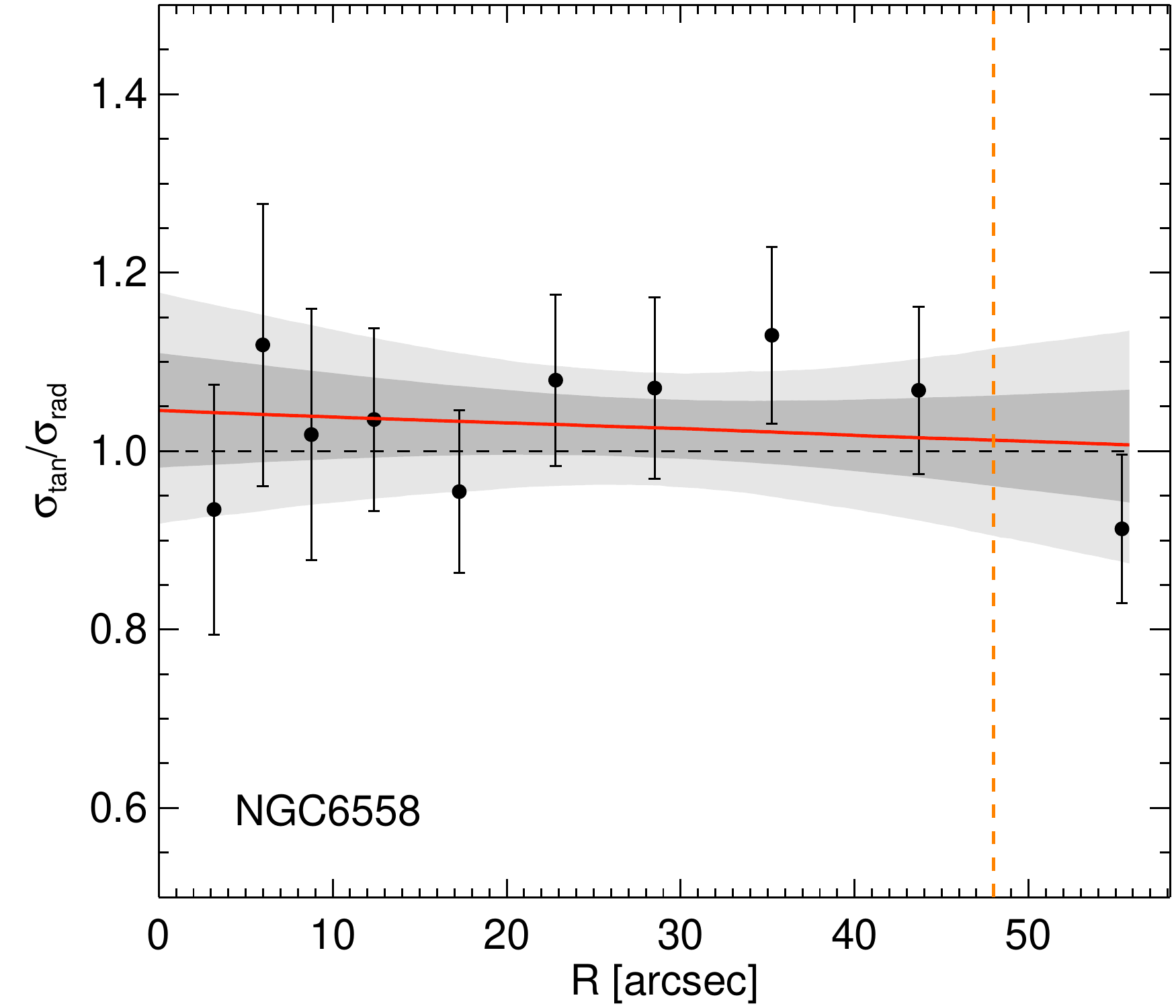}{0.33\textwidth}{}
	  \fig{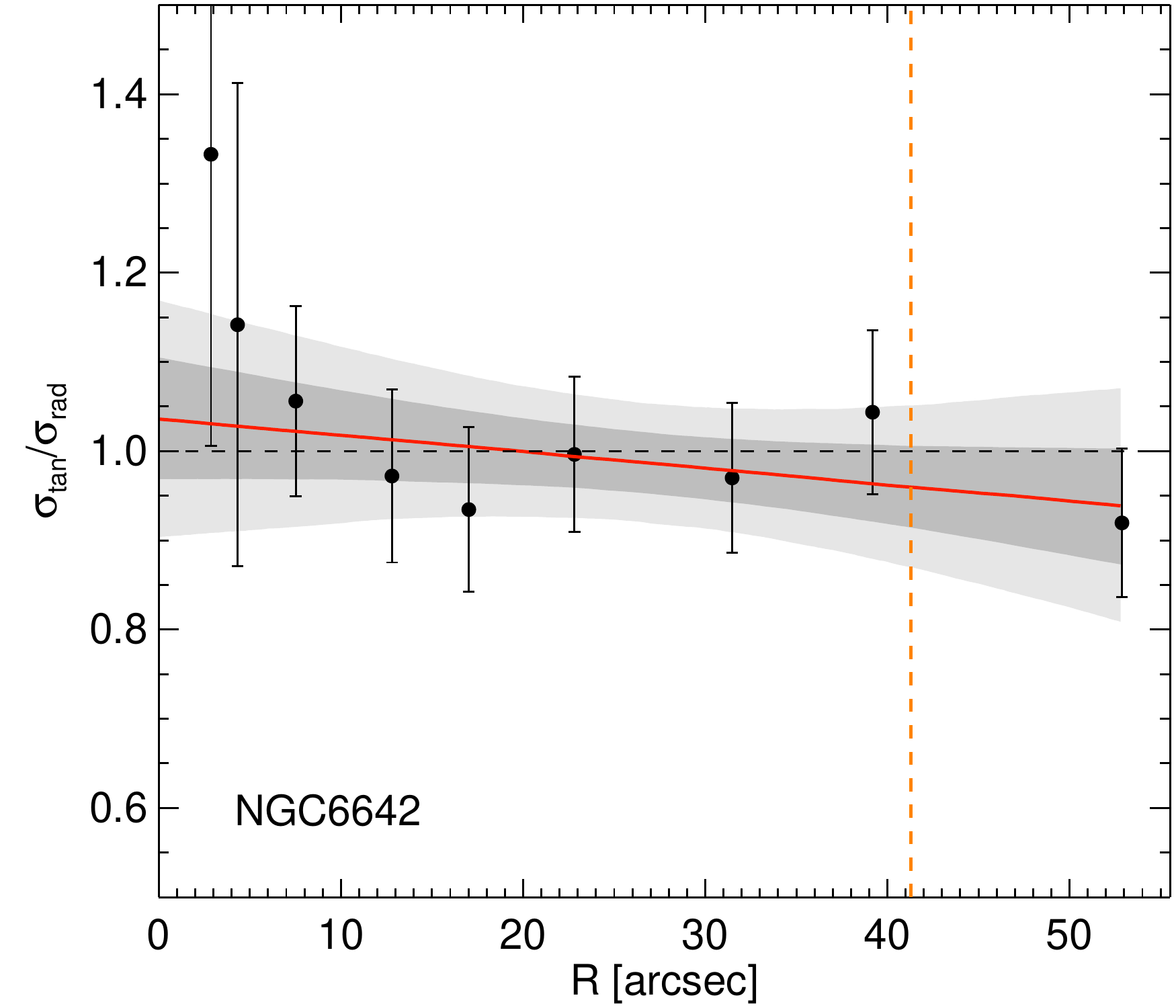}{0.33\textwidth}{}}
\caption{Anisotropy versus radius for our target clusters, defined as $\sigma_{\rm tan}$/$\sigma_{\rm rad}$ so that clusters with $\sigma_{\rm tan}$/$\sigma_{\rm rad}$$<$1 are radially anisotropic and $\sigma_{\rm tan}$/$\sigma_{\rm rad}$$>$1 are tangentially anisotropic, with isotropy indicated by the dashed horizontal line.  Linear fits are shown in red, and their 1-$\sigma$ (2-$\sigma$) uncertainty ranges are shown using dark (light) grey shading.  The location of $r_{\rm hl}$ is indicated using a dashed vertical orange line as in Fig.~\ref{vdisprfig}.
\label{anisofig}}
\end{figure}

We use our relative proper motions to characterize trends of anisotropy versus radius within our target clusters, where anisotropy is defined as the ratio of the tangential to radial proper motion dispersion $\sigma_{\rm tan} / \sigma_{\rm rad}$.  With this definition, isotropy is indicated by $\sigma_{\rm tan} / \sigma_{\rm rad}$=1, and values of $\sigma_{\rm tan} / \sigma_{\rm rad}$ larger (smaller) than unity indicate tangential (radial) anisotropy.  To measure anisotropy as a function of radius, we perform a linear fit following \citet{watkinskin}, again using \texttt{emcee}.  The best-fit straight lines are shown in red in Fig.~\ref{anisofig}, and in Table \ref{anisotab} we give the fit anisotropy at several characteristic radii for each cluster as in Table \ref{vdisptab}, and here we also include the linear gradient of $\sigma_{\rm tan} / \sigma_{\rm rad}$ with radius in the second column.  We find that our target clusters in general do not show statistically significant evidence for anisotropy, although we are only able to sample between 0.8 and 2.7$r_{\rm hl}$ depending on the cluster in question.  The only possible exceptions to this trend are NGC 6342, which appears radially anisotropic beyond $r_{\rm hl}$, and NGC 6380, which is the only cluster showing statistically significant tangential anistropy (also beyond $r_{\rm hl}$).  
In this context, we point out that while \citet{watkinskin} do not find any such clear incidence of tangential anisotropy in their sample, it may actually be expected on theoretical grounds.  In fact, N-body models predict that clusters orbiting in strong tidal fields should become tangentially anisotropic, at least in their outer parts \citep{bm03,vesperini14,sollima15,tiongco_aniso,bianchini17}, because tidal forces will, over time, preferentially remove stars on radial orbits.

\begin{deluxetable}{lccccc}
\tablecaption{Proper Motion Anisotropy at Selected Clustercentric Radii \label{anisotab}}
\tablehead{
\colhead{Cluster} & \colhead{$\delta(\sigma_{\rm tan}/\sigma_{\rm rad}) / \delta r$} & \colhead{$\sigma_{\rm tan}/\sigma_{\rm rad} (r=0)$} & \colhead{$\sigma_{\rm tan}/\sigma_{\rm rad} (r_{\rm 0})$} & \colhead{$\sigma_{\rm tan}/\sigma_{\rm rad} (0.5r_{\rm hl})$} & \colhead{$\sigma_{\rm tan}/\sigma_{\rm rad} (r_{\rm hl})$}  \\ & arcsec$^{-1}$ & & & & 
}
%%\colnumbers
\startdata
NGC6256 & $-0.001^{+0.001}_{-0.001}$ & $0.99^{+0.05}_{-0.05}$ & $0.98^{+0.04}_{-0.04}$ & $0.95^{+0.03}_{-0.03}$ &  \\
NGC6325 & $0.000^{+0.002}_{-0.002}$ & $0.98^{+0.07}_{-0.07}$ & $0.98^{+0.06}_{-0.06}$ & $0.98^{+0.05}_{-0.04}$ &  \\
NGC6342 & $-0.001^{+0.001}_{-0.001}$ & $0.98^{+0.05}_{-0.05}$ & $0.97^{+0.04}_{-0.04}$ & $0.95^{+0.03}_{-0.03}$ & $0.92^{+0.04}_{-0.04}$ \\
NGC6355 & $0.000^{+0.001}_{-0.002}$ & $0.95^{+0.06}_{-0.06}$ & $0.96^{+0.04}_{-0.04}$ & $0.96^{+0.03}_{-0.03}$ & $0.97^{+0.03}_{-0.03}$ \\
NGC6380 & $0.002^{+0.001}_{-0.001}$ & $0.94^{+0.03}_{-0.03}$ & $0.97^{+0.02}_{-0.02}$ & $0.99^{+0.01}_{-0.02}$ & $1.04^{+0.02}_{-0.02}$ \\
NGC6401 & $0.001^{+0.001}_{-0.001}$ & $1.00^{+0.03}_{-0.04}$ & $1.00^{+0.03}_{-0.03}$ & $1.01^{+0.02}_{-0.02}$ & $1.02^{+0.02}_{-0.02}$ \\
NGC6453 & $0.000^{+0.001}_{-0.001}$ & $1.01^{+0.05}_{-0.05}$ & $1.01^{+0.04}_{-0.04}$ & $1.01^{+0.02}_{-0.02}$ & $1.02^{+0.06}_{-0.05}$ \\
NGC6558 & $-0.001^{+0.002}_{-0.002}$ & $1.05^{+0.06}_{-0.06}$ & $1.04^{+0.05}_{-0.05}$ & $1.03^{+0.03}_{-0.03}$ & $1.01^{+0.05}_{-0.05}$ \\
NGC6642 & $-0.002^{+0.002}_{-0.002}$ & $1.04^{+0.07}_{-0.07}$ & $1.03^{+0.06}_{-0.06}$ & $1.00^{+0.04}_{-0.04}$ & $0.96^{+0.05}_{-0.05}$ \\
\enddata
%%%\tablecomments{Comments}
\end{deluxetable}

\subsection{Rotation}

Any rotation that is present in a given target cluster will be absorbed by the six-parameter linear transformations used to place stars in individual exposures onto a distortion-corrected master positional reference frame.  Since we measure proper motions relative to the bulk motion of the cluster, any non-cluster objects will then have a component imparted to their proper motion that is identical but opposite in sign as the true rotation of the cluster.  When a suitable background population (i.e.~with a proper motion that can be characterized in a straightforward way) is present, this aspect of the proper motion analysis can be exploited to measure cluster rotation \citep[e.g.][]{massari13,bellini_47tuc}.  While this is not the case for our target clusters, \citet{heyl47tuc} demonstrated that cluster rotation in the plane of the sky can be detected via skewness in the tangential proper motion distribution.  The skewness can be characterized by the $G_{\rm 1}$ value and its significance $Z_{\rm G1}$, where symmetric distributions have absolute values of $G_{\rm 1}$ close to zero, and the significance level $Z_{\rm G1}$ maps directly to $p$-values such that absolute values of $Z_{\rm G1}>$2 imply a detection at $>$2$\sigma$ \citep[e.g.][]{libralato_pm,bellini_pm}.  We list the values of $G_{\rm 1}$, $Z_{\rm G1}$, and (for convenience) the corresponding $p$-value in Table \ref{skewtab}.  We do not find evidence of rotation in any of the clusters in our sample, but only in two cases (NGC 6342 and NGC 6558) can we rule out rotation at a statistically significant ($\sim$2$\sigma$) level.  Such a lack of detectable rotation is perhaps expected on both observational and theoretical grounds: Simulations show that the radius from the cluster center where the rotation signal is strongest is at 1-2$r_{\rm hm}$ and decreases somewhat over time \citep{tiongco17}, while our high-quality proper motion sample only extends to $\lesssim$1.5$r_{\rm hm}$ in all but one case (and only extends to $<$$r_{\rm hm}$ for 3/9 target clusters).  On the observational side, the relative strength of rotation appears to anticorrelate with half-mass relaxation time \citep{sollima20}, rendering any rotational signal very difficult to detect given the likely old dynamical ages of our target clusters (Log $t_{\rm rh} \lesssim$9; see Sect.~\ref{relaxsect}).

\begin{deluxetable}{lccc}
\tablecaption{Skewness from Tangential Proper Motion Distribution \label{skewtab}}
\tablehead{
\colhead{Cluster} & \colhead{$G_{\rm 1}$} & \colhead{$Z_{\rm G1}$} & \colhead{$p$-value}
}
%%\colnumbers
\startdata
NGC6256 & -0.060 & -1.474 & 0.141 \\
NGC6325 & -0.020 & -0.275 & 0.783 \\
NGC6342 & -0.131 & -2.451 & 0.014 \\
NGC6355 & -0.003 & -0.066 & 0.947 \\
NGC6380 & 0.025 & 0.902 & 0.367 \\
NGC6401 & -0.018 & -0.750 & 0.453 \\
NGC6453 & -0.045 & -1.104 & 0.269 \\
NGC6558 & 0.095 & 1.967 & 0.049 \\
NGC6642 & 0.016 & 0.344 & 0.731 \\
\enddata
\end{deluxetable}

\section{Discussion: The Dynamical State of the Target Clusters \label{discusssect}}

With radial profiles of projected density, proper motion dispersion, and anisotropy in hand for our target clusters, we compare our sample to properties of Milky Way globular clusters at large.  Using three lines of reasoning, detailed below, we contend that our sample, comprised preferentially of clusters in the inner Milky Way, is highly dynamically evolved, and that the majority of our target clusters are either undergoing core collapse or are immediately pre- or post-core-collapse.

\subsection{Structural Parameters and King Profile Fits \label{densdiscusssect}}

The presence of a power-law (rather than flat) core in the density profile of a cluster indicates that it is in an advanced dynamical state.  Observationally, this is perhaps best supported by the power-law core of NGC 362 \citep{dalessandro362}, since its core-collapsed nature has been independently confirmed based on internal kinematics alone \citep{libralato_pm}.  In addition, \citet{ng06} showed that clusters with significant power-law slopes in their cores also tend to have shorter relaxation times (i.e.~they are dynamically old).  
On the theoretical side, simulations have consistently shown that cluster cores show a power-law slope when they are close to core collapse \citep[e.g.][]{chatterjee_cc,zocchi16}.  However, \citet{vesperini10} illustrate that even in the presence of larger sample sizes than we have at our disposal, observational errors render an assessment of whether a cluster is immediately pre- or post-core-collapse ambiguous.

While any clear detection of power-law cores in our target cluster density profiles is masked by observational uncertainties, we argue that six of our target clusters are immediately pre- or post-core-collapse, if not undergoing core collapse currently.
There are at least three (albeit not strictly independent) observational characteristics of core-collapsed cluster density profiles seen in N-body models reproduced by these six clusters:  First, our best-fitting \citet{king66} profiles underpredict the central density (with the exception of NGC 6642), in accord with simulations of clusters close to core collapse \citep{zocchi16}.  Second, all of these clusters have values of $r_{\rm c,obs}/r_{\rm hl} \approx$ 0.1, as found from density profile fits to simulated core-collapsed clusters \citep{chatterjee_cc}.  Third, all six of these clusters have best-fitting values of $W_{0} \gtrsim$8.5, or equivalently, $c \gtrsim$ 1.97 to within their uncertainties.  On the theoretical side, this is consistent with fits of \citet{king66} profiles to simulated clusters shortly before and after core collapse \citep{zocchi16}, and on the observational side, this is in line with the value of $W_{0}$=8.83 measured from the density profile of the core-collapsed cluster NGC 6752 \citep{ferraro6752}.  Lastly, we note that while the \citeauthor{h96} catalog lists two of our target clusters (NGC 6380 and NGC 6401) as candidate core-collapsed (and the rest as core-collapsed), \textit{all} of our target clusters were excluded from the surface brightness profile fits of \citet{mvdm} because the \citet{king66} fits by \citet{trager} yielded unreliable structural parameters.

\subsection{Dispersion and Concentration}

\begin{figure}
\gridline{\fig{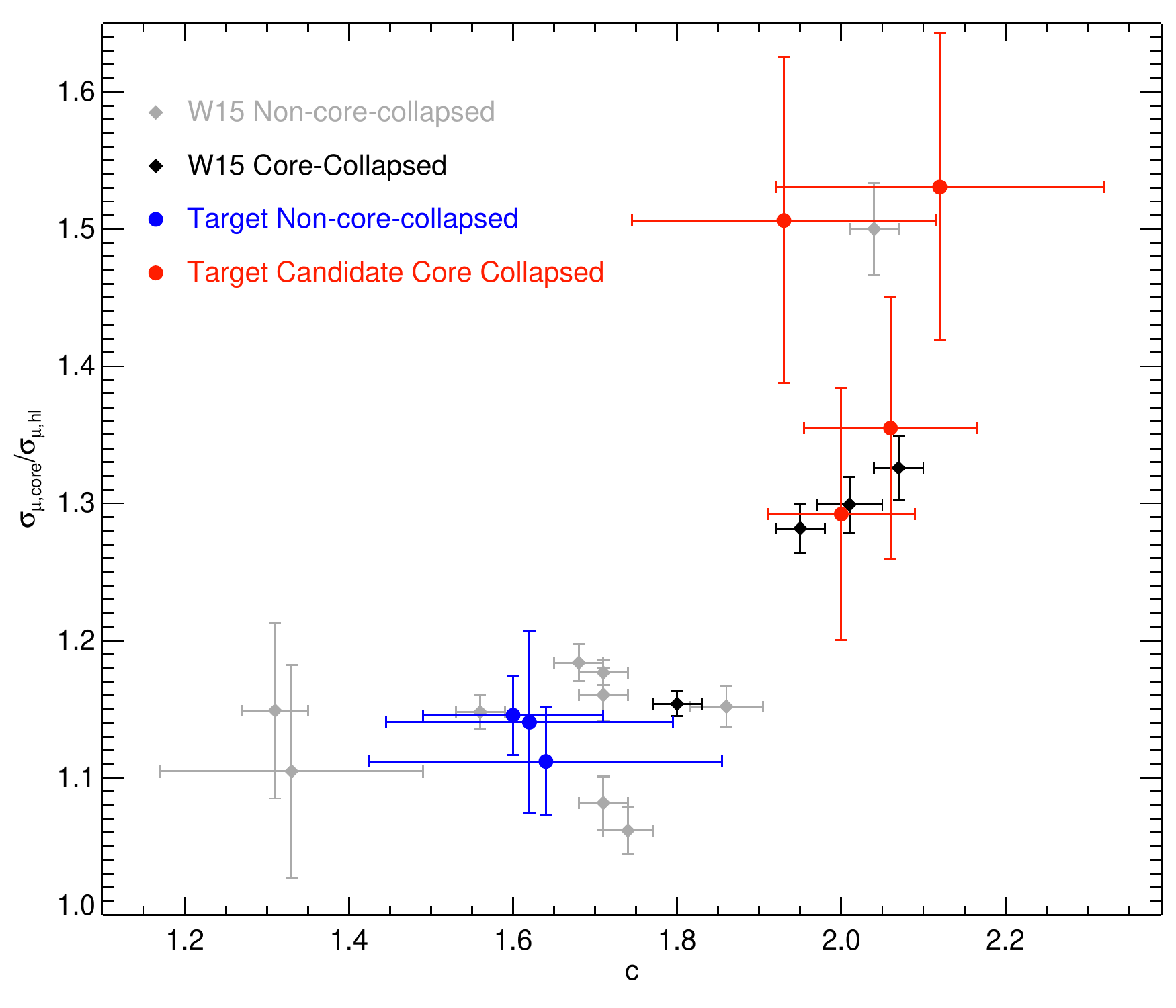}{0.9\textwidth}{}}
\caption{Ratio of the velocity dispersion at the King radius $r_{\rm 0}$ to velocity dispersion at the half-light radius $r_{\rm hl}$ as a function of concentration, after fig.~14 of \citet{watkinskin}.  Core-collapsed and non-core-collapsed clusters from \citet{watkinskin} are shown in black and grey respectively, and core-collapsed and non-core-collapsed clusters from our target sample are shown in red and blue respectively (NGC 6256 and NGC 6325 are not shown because their velocity dispersion profiles to not extend to $r_{\rm hl}$, but see text for further discussion).
\label{vdisp_concen_fig}}
\end{figure}

An empirical relationship between the steepness of the radial velocity dispersion profile and the cluster concentration was found by \citet{watkinskin}.  They use the ratio of the velocity dispersion at the King radius $r_{\rm 0}$ to the velocity dispersion at the half-light radius $r_{\rm hl}$ to characterize the steepness of the velocity dispersion slope as a function of distance from the cluster center, finding that clusters with higher concentrations have steeper dropoffs in their velocity dispersion profiles.  Their data are reproduced in Fig.~\ref{vdisp_concen_fig}, illustrating that core-collapsed clusters in their sample (shown in black) occupy a particular locus, whereas the remainder of the clusters (shown in grey) tend to have less steep dropoffs in their dispersion profiles.  In blue, we overplot our target clusters which are \textit{not} core-collapsed (NGC 6355, 6380, 6401), while the remainder of our sample, which we assert are close to core collapse, are shown in red.  This subset nicely coincides with the core-collapsed sample from \citet{watkinskin}, and shows even steeper dispersion slopes in a couple of cases (NGC 6453, 6558), albeit with larger uncertainties.

Although NGC 6256 and NGC 6325 are not shown in Fig.~\ref{vdisp_concen_fig} because their half-light radii are beyond the distance for which we were able to extract dispersion profiles, extrapolating their dispersion profiles out to $r_{\rm hl}$ yields values of $\sigma_{\mu}(r_{\rm 0})/\sigma_{\mu}(r_{\rm hl}) \simeq$ 1.6 and 1.4 respectively, which, when combined with their \citet{king66} concentrations in Table \ref{sbtab} places them with the other core-collapsed clusters in Fig.~\ref{vdisp_concen_fig}.  

\subsection{Anisotropy and Relaxation Time \label{relaxsect}}

Another empirical relationship characterized by \citet{watkinskin} is the correlation between the anisotropy (at the cluster core or half-light radii) versus relaxation time.  In particular, using relaxation times from the \citeauthor{h96} catalog, they found that regions in the clusters with relaxation times less than $\sim$0.55 Gyr (i.e.~the cores of their sample clusters) have had time to become isotropic, 
whereas most of the clusters in their sample, with longer half-light relaxation times, have not become relaxed out to their half-light radii.  Taking to heart their suggestion that this relation may be inverted as a means to constrain the relaxation times (i.e.~dynamical state) of other clusters, we use our linear fits from Sect.~\ref{anisosect} to calculate the anisotropy $\sigma_{\rm tan}/\sigma_{\rm rad}$ for our target clusters at their best-fit $r_{\rm 0}$ and $r_{\rm hl}$ from Table \ref{sbtab}.  From Table \ref{anisotab}, we find that \textit{all} of our target clusters have $\sigma_{\rm tan}/\sigma_{\rm rad}$=0.99 at their core radii to within their uncertainties, and the weighted mean (and its standard deviation) of $\sigma_{\rm tan}/\sigma_{\rm rad}$ at $r_{\rm 0}$ is 0.98$\pm$0.01, in good agreement with the results of \citet{watkinskin}.  At the half-light radii, however, we find a mean anistropy value across our sample of 1.01$\pm$0.01, or 1.00$\pm$0.01 if we extrapolate the linear fits of $\sigma_{\rm tan}/\sigma_{\rm rad}$ out to $r_{\rm hl}$ for NGC 6256 and NGC 6325.  This indicates that unlike the larger, more heterogenous sample of \citet{watkinskin}, our inner Milky Way sample is, on average, highly dynamically evolved.  Specifically, since our target clusters are isotropic even out to their half-light radii, the results of \citet[][see their sect.~5.4]{watkinskin} imply half-light relaxation times of Log $t_{\rm r,hl}$ $\lesssim$9\footnote{This range of values for $t_{\rm r,hl}$ is in good agreement with the half-mass relaxation times given for our target clusters by \citet{b19}, despite their possible overestimate of proper motion dispersions from \textit{Gaia} DR2 discussed in Sect.~\ref{vdispsect}.}.  Lastly, comparison with the core-collapsed cluster NGC 362 again provides independent support for the advanced dynamical status of our target clusters, this time via the anisotropy profile, as \citet{libralato_pm} found that NGC 362 members from the upper main sequence brightward are isotropic out to beyond $r_{\rm hl}$.

\section{Conclusions and Future Prospects \label{futuresect}}

We have calculated projected radial profiles of stellar density, proper motion dispersion, and anisotropy self-consistently for nine inner Milky Way GGCs, using MSTO-mass stars.  The projected profiles reveal that our sample is, in the mean, dynamically evolved, as revealed by their high concentrations, steep proper motion dispersion slopes, and lack of anisotropy out to our detection limits.  In particular, three of our target clusters (NGC 6355, NGC 6380, NGC 6401) are well-fit by single-mass \citet{king66} models, while the remaining six are likely currently undergoing (or very close to) core collapse, with central densities underpredicted by the King model fits and structural parameters consistent with other core-collapsed GGCs.

As we currently have at our disposal the most homogenous sample of dynamically old GGCs to date, additional epochs of well-dithered high-spatial-resolution imaging as we have obtained 
offer tantalizing possibilities to confirm or refute predictions of N-body models.  For example, the detection of tangential anisotropy in NGC 6380 bolsters predictions that given sufficient tidal forces, clusters will exhibit tangential anisotropy in their outer regions at late times.  However, these models make more specific testable predictions, namely that tidal forces are the critical driving factor determining the extent of this tangential anisotropy \citep{bianchini17} and that the tangential anisotropy is largely erased after core collapse \citep{bm03}.  Additional epochs of well-dithered imaging would also allow for proper motion dispersion measurements over a significant baseline of stellar mass to quantify the level of energy equipartion in our target clusters \citep[e.g.][]{libralato_pm,watkins20}, which our data quality unfortunately does not currently permit.  With this value in hand, the core-collapsed nature of any cluster can be independently verified using only internal kinematics \citep{libralato_pm,bianchini_concen}, while allowing a test of the predicted empirical relationship between concentration and mass segregation \citep{eta_concen_corr}, especially important in light of current discrepancies between theory and observations for two well-studied core-collapsed GGCs \citep{m15m30_mf}.  More generally, simulations predict that dynamically evolved ($t \gtrsim$10 $t_{\rm rc}$) clusters should have low mass-to-light ratios of M/L$\lesssim$3 
inside $r_{\rm hl}$ (and post-core-collapse dynamical evolution will act to lower the global mass-to-light ratio), and further predict that the exact values of M/L are metallicity dependent (\citealt{bianchini_ml}, but see \citealt{dalgleish_ml}).

Lastly, current and forthcoming near-IR massively multiplexed spectrographs and integral field units
will yield multi-element chemical abundances, but also precise radial velocities for large samples of cluster members.  The resulting velocity dispersion profiles, in physical units, may be combined with our proper motion dispersion profiles, in angular units, to estimate dynamical distances \citep{watkins_dyndist} which are independent of assumptions on the extinction law and its parameterization.  In fact, we have intentionally left our radial profiles in terms of angular rather than physical units precisely because the distances to our target clusters remain, for the most part, uncertain well beyond 10\%.  This is evidenced by both comparisons between optical and near-infrared photometric distances, (\citeauthor{h96},\citealt{valenti_cat1,valenti_cat2}) and substantial revisions to distances of bulge GGCs when extinction is properly accounted for and/or additional standard candles are available \citep[e.g.][]{javier_gcs,cadelano6256}.

\acknowledgements
It is a pleasure to thank the anonymous referee for a thorough reading of this manuscript and useful comments.  Support for program HST GO-15065 was provided by NASA through a grant from the
Space Telescope Science Institute, which is operated by the Association of
Universities for Research in Astronomy, Inc., under NASA contract NAS 5-26555. 

%\vspace{5mm}
\facilities{HST (ACS/WFC,WFC3/UVIS)}

%\software{IRAF, IDL}

%\appendix

%Appendix goes here.

\end{document}